\documentclass[preprint,amsmath,amssymb,aps,prd,showpacs]{revtex4-1}
 \pdfoutput=1

\usepackage{float}

\usepackage{epsfig}
\usepackage{dcolumn}
\usepackage{bm}
\usepackage{tikz}
\usepackage{graphicx, graphics, color}
\usepackage{dcolumn}
\usepackage{bm}
\usepackage{hyperref}
\usepackage[mathlines]{lineno}
\usepackage{float}
\usepackage{subfig}

\newcommand{\be}{\begin{equation}}
\newcommand{\ee}{\end{equation}}
\newcommand{\ben}{\begin{eqnarray}}
\newcommand{\een}{\end{eqnarray}}
\newcommand{\la}{{\lambda}}

\newcommand{\cL}{{\cal L}}

\newcommand{\cE}{{\cal E}}

\newcommand{\p}{\partial}
\newcommand{\na}{\nabla}

\newcommand{\tpe}{{\tilde p}}

\newcommand{\tA}{\tilde A}
\newcommand{\tF}{\tilde F}

\newcommand{\tD}{\tilde D}

\newcommand{\ep}{\epsilon}
\newcommand{\tep}{\tilde \epsilon}

\newcommand{\ga}{\gamma}

\newcommand{\tB}{{\tilde B}}

\newcommand{\tbeta}{{\tilde \beta}}

\newcommand{\te}{\tilde e}

\graphicspath{{figs/}}
\graphicspath{{../figs/}}

\begin{document} 

\title{ Influence of dark photon on magnetized and charged particle orbits around static spherically symmetric black hole}

\author{Marek Rogatko} 
\email{rogat@kft.umcs.lublin.pl}
\author{Paritosh Verma} 
\email{pverma@kft.umcs.lublin.pl}
\affiliation{Institute of Physics, 
Maria Curie-Sklodowska University, 
pl.~Marii Curie-Sklodowskiej 1,  20-031 Lublin,  Poland}


\date{\today}

\begin{abstract}
We elaborate the problem of magnetized
particle motion in the spacetime of a static, spherically symmetric black hole
influenced by weak magnetic fields stemming from visible and dark matter sectors.
The Wald's procedure for obtaining the weakly magnetized solution, generalized to the case of dark photon - Einstein-Maxwell gravity was implemented.
The collision process analysis of two particles in the background of the black hole has been studied in order to find the signature of dark matter presence in the nearby of the object in question.
\end{abstract}

\maketitle
\flushbottom

\section{Introduction}
\label{sec:intro}



It is evident from various astronomical observations that the visible matter in our Universe accounts for only about 4.9 \% of its total mass and 95.1 \% is still unknown to us
\cite{planck}.
The
unknown component is referred as the {\it dark sector} of the Universe and divides it into two parts: {\it dark matter} and dark energy. {\it Dark matter}  constitutes of 26.6 \% of the total mass and it influences among all on galaxy rotation curves, measurements of cosmic microwave background radiation,
baryonic oscillations \cite{ber18a,ber18b}. Pulsar timing array experiments \cite{sma23} have also revealed its impact.

Although we have indirect evidence of {\it dark sectors}, neither we have detected them experimentally nor we have a concrete theory to explain them. The Standard Model, fails to explain the nature of the {\it invisible sector}. However
one of the simplest way out of this problem is to suppose 
 the existence of particles weakly interacting with the ordinary matter.

A very plausible candidate for explaining physics beyond the Standard Model, is the concept of {\it dark photon}, being hypothetical Abelian gauge boson coupled to the
ordinary Maxwell gauge field \cite{hol86}.
The idea in question acquires also justification in
the contemporary unification scheme  \cite{ach16}, where the mixing portals coupling Maxwell 
and auxiliary gauge fields charged under their own groups are under inspection. 

On the other hand, it has been claimed that {\it dark photons} might be produced during inflationary fluctuations \cite{gra16,sat22}, axion oscillations \cite{axionosc},
 during exotic particles decays \cite{dro19}, eheating \cite{reheat}, and from cosmic strings \cite{cstrings}.

Several anomalous astrophysical effects like $511~ keV$
gamma rays \cite{jea03}, excess of the positron cosmic ray flux in galaxies \cite{cha08}, and the observations of an anomalous monochromatic $3.56~ keV$ X-ray line in the spectrum of some galaxy clusters \cite{bub14}, may justify the aforementioned {\it dark photon} idea.
Many astrophysical and laboratory experiments \cite{fil20}, 
are dedicated to establish the range of values for {\it dark photon} - Maxwell field coupling constant and the mass of the hidden photon.
Among all they are connected with the
studies of gamma rays emissions from dwarf galaxies \cite{ger15},
inspection of dilaton-like coupling to photons caused by ultra-light {\it dark matter} \cite{bod15} 
 fine structure constant oscillations \cite{til15}, {\it dark photon} emission during supernova event \cite{cha17}, 
 electron excitation measurements in CCD-like detector \cite{sensei}, the search for a {\it dark photon} in $e^+ e^-$ collisions at BABAR experiment \cite{lee14},
 measurements of the muon anomalous effect \cite{dav11}.
 
Recently it has been found the new exclusion limit for the $\alpha$-coupling constant $\alpha =1.6 \times
10^{-9}$ and the mass range of {\it dark photon}
$2.1\times 10^{-7} - 5.7 \times10^{-6} eV$ \cite{rom23} -\cite{fil23}, while
using quantum limited amplification,  the first probing of the kinetic mixing coupling constant to $10^{-12}$ level for majority of {\it dark photon} masses was given in \cite{ram23}. 
An upper bound on coupling constant  $\alpha < 0.3-2 \times 10^{-10}$ (at 95 percent confidence level)  has been revealed in Ref. \cite{kot23}.

In the context of the absence of the evidence of the most popular {\it dark matter} candidates in experimental data, the diversifying of the quest efforts
for {\it dark sector} should be in order \cite{ber18b},
taking into account astronomical surveys, gravitational wave observatories and new concepts of Earth experiments.
 One of the most promising way of finding the manifestation of {\it dark matter} presence seems to be studies of the environments of compact objects like black holes,
 wormholes and compact star-like objects. For instance some problems connecting with {\it dark matter} clouds and their influence on compact objects were treated in 
 \cite{kic19}-\cite{kic22}, while the uniqueness of black hole in {\it dark photon} Einstein-Maxwell gravity was analyzed in \cite{rog23,rog24}.

 In the last few years, there have been also various attempts to study the motion of particles around various classes of black holes and wormholes. 
Namely if one restricts to the case of static black hole representative samples, authors have worked on spinning particles motion in background of Schwarzschild  black holes \cite{abd23},
 hairy black holes in Horndeski gravity \cite{ray23}, rotating \cite{abd22} and traversable wormholes \cite{ben21}, as well as,
 Schwarzschild-MOG black holes  spacetime \cite{bob23}. 
On the other hand, the problem of both charged \cite{fro10}-\cite{,zah13} and magnetized \cite{fel03} particle motion in the weakly magnetized Schwarzschild,
non-Schwarzschild \cite{ray16}, and Gauss-Bonnet  \cite{abd20} black holes spacetime was widely treated.

The main aim of our work will be to look for the potential new effects caused by {\it dark sector} via studies of magnetized/ charged
particle motions around static black hole. The black hole will be weakly magnetized by Maxwell and {\it dark photon}
magnetic fields. As a {\it dark sector} we consider auxiliary $U(1)$-gauge field coupled to the ordinary Maxwell one.
Especially we pay attention to the influence of the {\it dark photon} field magnitude, as well as, the coupling constant between {\it visible } and {\it dark sector} fields,
on the discussed phenomena. In the case of coupling constant impact we shall mainly look for quantitative behaviors, due to its small value.


The organization of the paper is as follows. In Sec. \ref{sec:darkphoton} we present the model  of {\it dark photon}
and find the generalized four-momentum of massive particle moving in two $U(1)$-gauge fields. In Sec. III one describes the weakly magnetized
 black hole solution achieved by the generalization of the Wald's method \cite{wal74}
 to the case of {\it visible}
 and {\it dark matter} $U(1)$-gauge fields. Sec. IV will be devoted to studies of magnetized particle motion around black hole from Sec. II. One uses
 Hamilton-Jacobi equations and obtains the magnetic coupling parameter, showing that magnetized particle in the vicinity
 of black hole weakly magnetized by {\it visible}
 and {\it hidden sector} fields has energy and angular momentum dependent on them.
 In Sec. V we shall pay attention to the collision problem of two particles moving in the equatorial orbit around weakly magnetized by {\it visible}
 and {\it hidden} $U(1)$-gauge fields. One considers the case of two magnetized, magnetized and charged, as well as, neutral particles.
Sec. VI is connected with the charged particle motion in homogeneous, dipolar, and parabolic magnetic field. We shall look for the influence of {\it dark sector}
on the motion in question. In Sec. VII we conclude our investigations.

\section{The Model of dark photon} \label{sec:darkphoton}
In this section we shortly describe the {\it dark photon} model and propose a transformation of underlying gauge fields in order to simply 
the underlying action, i.e., dispose of {\it kinetic mixing} term. In the process of this we obtain new gauge fields, being the mixture of the starting ones, with adequate factors
comprising $\alpha$-coupling constant. Next, one derives the equation of motion for charged particle affected by two gauge fields.

The action describing two coupled, massless gauge field is given by
\be
S_{M-dark~ photon} = \int  d^4x  \Big(
- F_{\mu \nu} F^{\mu \nu} - B_{\mu \nu} B^{\mu \nu} - {\alpha}F_{\mu \nu} B^{\mu \nu}
\Big),
\label{ac dm}
\ee  
where $\alpha$ is taken as a coupling constant.

 To get rid of the {\it kinetic mixing } term we define new gauge fields. They can be written  as 
\ben \label{transA}
\tA_\mu &=& \frac{\sqrt{2 -\alpha}}{2} \Big( A_\mu - B_\mu \Big),\\ \label{transB}
\tB_\mu &=& \frac{\sqrt{2 + \alpha}}{2} \Big( A_\mu + B_\mu \Big).
\een
As a result one leaves with only modified gauge fields, i.e., 
\be
 F_{\mu \nu} F^{\mu \nu} +
B_{\mu \nu} B^{\mu \nu} +  \alpha F_{\mu \nu} B^{\mu \nu}
\Longrightarrow
 \tF_{\mu \nu} \tF^{\mu \nu} +
\tB_{\mu \nu} \tB^{\mu \nu},
\ee
where we set $\tF_{\mu \nu} = 2 \na_{[\mu }\tA_{\nu ]}$, and respectively $\tB_{\mu \nu} = 2 \na_{[\mu }\tB_{\nu ]}$. Just the action can be rewritten as
\be
S_{M-dark~ photon}  = \int d^4x  \Big(
- \tF_{\mu \nu} \tF^{\mu \nu} - \tB_{\mu \nu} \tB^{\mu \nu}
\Big).
\label{vdc}
\ee
Variation of the action (\ref{vdc}) with respect to $\tA_\mu$ and $\tB_\mu$ reveals the following equations of motion for Maxwell {\it dark matter} system:
\be
\na_{\mu} \tF^{\mu \nu } = 0, \qquad \na_{\mu} \tB^{\mu \nu } = 0.
\label{fb}
\ee

 According to the relations (\ref{vdc}) and (\ref{fb}), we shall search for the equation of motion of charged particle in the background of the modified gauge fields. 
 However, the consistency of the approach requires the appropriate redefinitions of the charges coupled to original fields.
 The  resulting action of massive charged particle influenced by both visible and {\it dark matter} sectors is assumed to be 
\be
S = - \int m \sqrt{- ds^2} + \te_A \int \tA_{\mu} dx^\mu + \te_B \int \tB_\mu dx^\mu,
\ee
where  for transformed charges are defined as
\ben \label{cA}
\te_A &=& \frac{\sqrt{2 -\alpha}}{2} \Big( e - e_d \Big),\\ \label{cB}
\te_B &=& \frac{\sqrt{2 + \alpha}}{2} \Big( e + e_d \Big).
\een
In the above equations $e$ stands for the Maxwell charge, while $e_d$ is connected with {\it dark sector} one. 
In our considerations we assume that the charges $e$ and $e_d$ (as well as $\te_A $ and $\te_B$) are positive.

The standard calculation leads to the following equation of motion
\be
m~\frac{D u^\mu}{d \tau} = \Big( \te_A \tF^{\mu \nu} +  \te_B \tB^{\mu \nu} \Big) u_\nu,
\label{eqmot}
\ee
where $\frac{D}{d \tau}$ denotes in general case the covariant derivative with respect to the proper time.

As a result the four-momentum of the massive particle subject to two gauge fields 
may be written as
\be
\tpe_\mu = m u_\mu + \te_A \tA_\mu +  \te_B  \tB_\mu.
\label{mom}
\ee
We can notice that transforming the charges and fields back one arrives at 
\be
\tpe_\mu = m u_\mu + e A_\mu + e_d B_\mu  + \frac{\alpha}{2} \Big( e_d A_\mu + e B_\mu \Big).
\label{cons}
\ee
Having in mind the action (\ref{ac dm}), 
the above equation (\ref{cons}) provides the consistency in choosing transformed charges (\ref{cA})-(\ref{cB}) and fields (\ref{transA}) and (\ref{transB}).

It can be remarked that for the special case, when $e_d=0$, the above relation reduces to the following:
\be
\tpe_\mu = m u_\mu + e \Big(A_\mu + \frac{\alpha}{2}B_\mu \Big).
\ee
This shows the modifications of the kinetic momentum by dark sector gauge field.

\section{Weakly magnetized black holes}

In Ref. \cite{wal74} it has been revealed that $A_\mu$ potential can be defined as a linear combination of Killing vector fields, due to the fact that in vacuum spacetime, Killing vectors
satisfy the similar equation as $A_\mu$ in Lorenz gauge.

Returning to the case under consideration, it can be observed that the equations (\ref{fb}), in Lorenz gauge, written for spherically symmetric solution of Einstein equation with the condition of vanishing Ricci tensor $R_{\mu \nu}$, reveal
 that $\square \tA^\mu = 0$ and $\square \tB^\mu = 0$, where $\square = g^{\mu \nu} \na_\mu \na_\nu$. 
 Then, using the fact that any Killing vector can be composed as a linear combination of the other ones,
 we define the potentials $\tA_\mu$ and $\tB_\mu$ in the forms as follows:
\ben
\tA^\mu &=& \ga_1~ \xi^\mu_{(t)} + \ga_2 ~\xi^\mu_{(\phi)},\\ 
\tB^\mu &=&  \delta_1~ \xi^\mu_{(t)} + \delta_2 ~\xi^\mu_{(\phi)}.
\een
Consequently, the Killing vector fields $\xi^\mu_{(t)}$ and $\xi^\mu_{(\phi)}$ generate time translation and rotations around the symmetry axis.

In the next step, having in mind definitions of ADM mass, angular momentum and charges
\ben
- 8 \pi M &=& \int \ep_{\alpha \beta \ga \delta} \na^\ga \xi^\delta_{(t)}, \qquad
16 \pi J = \int \ep_{\alpha \beta \ga \delta} \na^\ga \xi^\delta_{(\phi)},\\ \label{charges}
4 \pi Q(\tF) &=& \int \ep_{\alpha \beta \ga \delta} \tF^{\ga \delta}, \qquad 
4 \pi Q(\tB) = \int \ep_{\alpha \beta \ga \delta} \tB^{\ga \delta},
\een
it can be revealed that the aforementioned $U(1)$-gauge potentials are given by
\ben
\tA^\mu &=& \frac{B^{(\tF)}_0}{2} \Big[ \xi^\mu_{(\phi)} + 2~a \xi^\mu_{(t)} \Big] - \frac{Q(\tF)}{2 M} \xi^\mu_{(t)},\\
\tB^\mu &=& \frac{B^{(\tB)}_0}{2} \Big[ \xi^\mu_{(\phi)} + 2~a \xi^\mu_{(t)}  \Big] - \frac{Q(\tB)}{2 M} \xi^\mu_{(t)},
\een
where we have denoted $a = J/M$. For the consistency with the later considerations, we set that the mixture of magnetic fields components connected with {\it visible} and {\it dark sectors} implies
\be
B^{(\tF)}_0 = \frac{\sqrt{2 -\alpha}}{2} \Big( B^{(F)}_0 - B^{(B)}_0 \Big),
\qquad
B^{(\tB)}_0 = \frac{\sqrt{2 + \alpha}}{2} \Big( B^{(F)}_0 + B^{(B)}_0 \Big).
\label{bbb}
\ee
On the other hand, from the definition of the charges (\ref{charges}), one has that the following relations are provided:
\be
Q{(\tF)} = \frac{\sqrt{2 -\alpha}}{2} \Big( Q{(F)} - Q{(B)} \Big),
\qquad
Q{(\tB)} = \frac{\sqrt{2 + \alpha}}{2} \Big( Q{(F)} + Q{(B)} \Big),
\ee
where the charges $Q(F)$ and $Q(B)$ are defined as in the equation (\ref{charges}), for Maxwell and {\it dark photon}, respectively.

In the case when $a = J/M =0$, one obtains the static black hole case. Moreover if $Q=0$
 the mixture of gauge field components is given by
 \be 
 \tA^\mu = \frac{ B^{(\tF)}_0}{2} \xi^\mu_{(\phi)}, \qquad  \tB^\mu = \frac{B^{(\tB)}_0}{2} \xi^\mu_{(\phi)}.
\label{stat}
\ee
Having in mind the relation (\ref{bbb}), one can obtain that the adequate components of {\it visible} and {\it dark} sectors four-potentials. They are written as
 \be 
 A^\mu = \frac{ B^{(F)}_0}{2} \xi^\mu_{(\phi)}, \qquad  B^\mu = \frac{B^{(B)}_0}{2} \xi^\mu_{(\phi)}.
\label{stat1}
\ee

Further,
it can be shown that the magnetic fields seen
by a local observer, being at rest in the spacetime in question, yield
\ben
\tB^{(\tF) \mu} &=& B^{(\tF)}_0 \sqrt{f(r)} \Big( \cos \theta ~\delta^\mu_r - \frac{1}{r} \sin \theta ~\delta ^\mu_\theta \Big),\\
\tB^{(\tB) \mu} &=& B^{(\tB)}_0 \sqrt{f(r)} \Big( \cos \theta ~\delta^\mu_r - \frac{1}{r} \sin \theta ~\delta ^\mu_\theta \Big).
\een
In the terms of $U(1)$-gauge field components, due to the relations (\ref{bbb}), they are provided by
\ben
B^{(F) \mu} &=& B^{(F)}_0 \sqrt{f(r)} \Big( \cos \theta ~\delta^\mu_r - \frac{1}{r} \sin \theta ~\delta ^\mu_\theta \Big),\\
B^{(B) \mu} &=& B^{(B)}_0 \sqrt{f(r)} \Big( \cos \theta ~\delta^\mu_r - \frac{1}{r} \sin \theta ~\delta ^\mu_\theta \Big).
\een

\section{Hamilton-Jacobi Equations}
In our studies we shall pay attention to the problem of motion of magnetized  particle around static spherically symmetric black hole, taking into account the external magnetic fields stemming from
{\it visible} and {\it dark sectors}.

It leads to the following form of Hamilton-Jacobi equations
\be
g^{\mu \nu} p_\mu p_\nu = - \Big(
m - \frac{1}{2} \tD^{\mu \nu}_{(\tF)} ~\tF_{\mu \nu}  -  \frac{1}{2} \tD^{\mu \nu}_{(\tB)} ~ \tB_{\mu \nu} \Big)^2.
\ee
 The scalar products emerging of the right-hand side of Hamilton-Jacobi equation will be responsiblefor the interactions of magnetized particle with {\it visible } and {\it dark} magnetic fields. Namely,
the antisymmetric tensor $\tD^{\mu \nu}_{(\tF)~(\tB)} $ describes properties of the particle immersed in both $U(1)$-gauge fields. Their explicit forms are provided by
\ben \label{mv}
\tD^{\mu \nu}_{(\tF)} ~\tF_{\mu \nu} &=& \frac{\sqrt{2 -\alpha}}{2} \Big( \mu^{(F)}_\delta - \mu^{(B)}_\delta \Big)~\tep^{\mu \nu \rho \delta} u_\rho~ \tF_{\mu \nu},\\ \label{md}
\tD^{\mu \nu}_{(\tB)} ~ \tB_{\mu \nu} &=& \frac{\sqrt{2 +\alpha}}{2} \Big( \mu^{(F)}_\delta + \mu^{(B)}_\delta \Big)~\tep^{\mu \nu \rho \delta} u_\rho~ \tB_{\mu \nu},
\een
which are in accord with the consistency of the {\it dark photon} theory.
$\mu^{(F)}_\delta$ and $\mu^{(B)}_\delta $ stand for the four-vectors of dipole moments of magnetic particle bounded with Maxwell and {\it dark photon} fields, respectively. 
$\tep_{\mu \nu \alpha \beta}$ stands for 
the pseudo-tensorial density, for the considered 
spacetime with the determinant of metric tensor $g$.

The Hamilton-Jacobi equations take the forms as follows:
\ben
\label{HJ-eq}
g^{\alpha \beta}~\Big( \frac{\p S}{\p x^\alpha} &-& \te_A \tA_\alpha - \te_B \tB_\alpha \Big) 
\Big( \frac{\p S}{\p x^\beta} - \te_A \tA_\beta - \te_B \tB_\beta \Big) \\ \nonumber
&=& - m^2 + m~\tD^{\mu \nu}_{(\tF)} ~\tF_{\mu \nu} + m~\tD^{\mu \nu}_{(\tB)} ~ \tB_{\mu \nu} ,
\een
where $\frac{\p S}{\p x^\mu} = \tpe_\mu$. In the above relation
one takes into account the presence of the weakly magnetized static spherically symmetric black hole, which
enables us to suppose that quadratic-type terms of the form tend to zero, i.e.,
$\Big( \tD^{\mu \nu}_{(\tF)~(\tB)} \tF_{\mu \nu} (\tB_{\mu \nu} )\Big)^2 
\rightarrow 0$. 

In the case when $e_d=0$ and $\mu^\delta_{(B)} =0$ the Hamilton-Jacobi relations reduce to the following:
\ben
g^{\mu \nu}~\Big[ \frac{\p S}{\p x^\mu} &-& e \Big( A_\mu + \frac{\alpha}{2} B_\mu \Big) \Big]
\Big[ \frac{\p S}{\p x^\nu} - e \Big( A_\nu + \frac{\alpha}{2} B_\nu \Big) \Big] \\ \nonumber
&=&  - m^2 + m~\Big[ \tep^{\mu \nu \rho \delta} \mu_\delta^{(F)} u_\rho ~\Big( F_{\mu \nu} + \frac{\alpha}{2} B_{\mu \nu} \Big) \Big].
\een
The above relation constitutes the generalize the equation derived in Ref. \cite{fel03} (equation (31)) for the Maxwell electrodynamic case, and in the case of $B_\mu =0,~\alpha =0$
reduces to the aforementioned one.

To proceed further, let us rewrite field strengths tensor in terms of their electric and magnetic components. Namely one has that
\ben \label{ff}
\tF_{\mu \nu} &=& 2 u_{[\mu} E_{\nu ]}^{(\tF)} - \tep_{\mu \nu \alpha \beta}~ u^\alpha B^{\beta (\tF)}, \\ \label{bb}
\tB_{\mu \nu} &=& 2 u_{[\mu} E_{\nu ]}^{(\tB)} - \tep_{\mu \nu \alpha \beta} ~u^\alpha B^{\beta (\tB)},
\een
where we set for the electric components 
\be
E_a^{(\tF)} = - \tF_{0a}, \qquad E_a^{(\tB)} = - \tB_{0a},
\ee
and for magnetic ones
\be
B_a^{(\tF)} = \frac{1}{2} \ep_{abc} \tF^{bc}, \qquad B_a^{(\tB)} = \frac{1}{2} \ep_{abc} \tB^{bc}.
\ee
Relations (\ref{ff}) and (\ref{bb}), enable us to rewrite the action of the antisymmetric tensor $\tD^{\mu \nu}$ on the adequate field strengths, in the form as
\be 
\tD^{\mu \nu}_{(\tF)} ~\tF_{\mu \nu} + \tD^{\mu \nu}_{(\tB)} ~ \tB_{\mu \nu} =
2 \Big[ \mu_\ga^{(F)}  \Big( B^{\ga (F)} +    \frac{\alpha}{2} B^{\ga (B)} \Big)  
+ \mu_\ga^{(B)}  \Big( B^{\ga (B)} +    \frac{\alpha}{2} B^{\ga (F)} \Big)\Big) \Big].
\label{dfb}
\ee
It can be seen that the above relation reduces to the form of
$2 \mu_\ga^{(F)} \Big( B^{\ga (F)} + \frac{\alpha}{2} B^{\ga (B)} \Big)$, when we suppose that $\mu_\ga^{(B)}$ is equal to zero.

As in Ref. \cite{fel03}, we shall use the projection of $\mu^{(F), (B)}_\delta$ and the $B^{(F),(B)}_\delta$ on an orthonormal tetrad frame, adopted 
for the fiducial observer $\la_{\hat 0} = u$. In what follows the hatted index will refer to the tetrad frame.
Consequently the right-hand side of relation \ref{dfb}) is equal to
\be
\tD^{\mu \nu}_{(\tF)} ~\tF_{\mu \nu} + \tD^{\mu \nu}_{(\tB)} ~ \tB_{\mu \nu} =
2 \cL(\la_{\hat 0}) \Big[
\mu^{(F)} B_0^{(F)} + \mu^{(B)} B_0^{(B)} + \frac{\alpha}{2} \Big( \mu^{(F)} B_0^{(B)} + \mu^{(B)} B_0^{(F)} \Big) \Big],
\ee
where we have denoted the modulus of the adequate magnetic moment of the particle in question, with respect to the {\it visible} and {\it dark sector} fields
\be
\mu^{(F)} = \Big( \mu^{(F) \delta} \mu^{(F)}_\delta \Big)^{\frac{1}{2}}, \qquad
\mu^{(B)} = \Big( \mu^{(B) \delta} \mu^{(B)}_\delta \Big)^{\frac{1}{2}}, 
\ee
while $\cL(\la_{\hat 0}) $ describes the nature of comoving reference frame with the studied particle rotating around the compact object. $\cL(\la_{\hat 0})$ is a function depending on spacetime 
coordinates and the parameters of the defined tetrad, connected with the observer \cite{fel03}.

As we study the circular motion of a magnetized particle moving around weakly magnetized static spherically symmetric black hole
\be
ds^2 = - f(r)dt^2 + g(r) dr^2 + r^2 \Big( \sin^2 \theta d\phi^2 + d \theta^2 \Big),
\ee
taking into account
the influence fo {\it dark sector},
the symmetry of the problem (circular motion in the equatorial plane $\theta = \frac{\pi}{2}$, perpendicularity of magnetic moment to this plane with non-zero component
$\mu^\theta$, the axial configuration
of magnetic field) enable us to
write
 the action for the problem in question as follows:
\be
\label{sol-HJ}
S = - Et + L \phi + S_r(r),
\ee
where $p_t = -E,~p_\phi = L$.
 
 It leads to the following result:
\be
\Big( \frac{dr}{d \tau} \Big)^2 = \frac{\cE^2}{f(r)~g(r)} - 1 - 2 U_{eff} (f(r), g(r), l, \beta, \cE),
\label{dif r}
\ee
where $\tau$ is the proper time, and the effective potential is defined as
\be
U_{eff} = \frac{1}{2} \Bigg[ \frac{1}{g(r)} \Big[
\Big( 1 + \frac{l^2}{r^2} \Big) - \Big( \beta(F) + \beta(B) + \frac{\alpha}{2}  (\beta(F,B) + \beta(B,F)) \Big) \cL(\la_0) \Big] -1 \Bigg],
\ee
while its components are provided by
\be
\cE = \frac{E}{m}, \qquad l = \frac{L}{m},
\ee
for specific energy and angular momentum of the particle and coefficients $\beta$
\ben
\beta(F) &=& \frac{2 \mu^{(F)} B_0^{(F)}}{m},  \qquad \beta(B)= \frac{2 \mu^{(F)} B_0^{(F)}}{m}, \\
\beta(F, B) &=&  \frac{2 \mu^{(F)} B_0^{(B)}}{m},  \qquad \beta(B,F) =  \frac{2 \mu^{(B)} B_0^{(F)}}{m}.
\een
We have denoted 
by $\beta$ the magnetic coupling parameters illustrating the strengths of the adequate external magnetic effects caused by $U(1)$-gauge fields
pertaining to {\it visible } and {\it dark matter} sectors.

Further, let us proceed to find the conditions for circular orbits around static spherically symmetric black hole.
Namely, one has
\be
 \frac{dr}{d \tau} = 0, \qquad \frac{\p U_{eff}}{\p r} = 0.
 \label{circ}
 \ee
From the first relation of (\ref{circ}) one can establish the following equation for the parameters $\beta$
\ben \label{bbd}
\tbeta(F,B,\alpha) &=& \Big( \beta(F) + \beta(B) + \frac{\alpha}{2}  (\beta(F,B) + \beta(B,F)) \Big) \\ \nonumber
&=& \frac{1}{\cL(\la_{\hat 0})} g(r)~\Big[ \frac{1}{g(r)} \Big( 1 + \frac{l^2}{r^2} \Big) - \frac{\cE^2}{f(r) g(r)} \Big].
\een
On the other hand, the second relation of (\ref{circ})
and equation (\ref{dif r}) reveal that the partial derivative of the effective potential can be expressed as
\be
\frac{\p U_{eff}}{\p r} = - \frac{g'(r)~\cE^2}{g^2(r) f(r)} + \frac{1}{g(r)} \Big( -\frac{2 l^2}{r^3} - \tbeta(r) ~\cL'(\la_{\hat 0}) \Big),
\ee
which under the condition that $f(r)~g(r) =1$ implies the following:
\be
\frac{\p U_{eff}}{\p r} = f(r)~\cL(\la_{\hat 0})~\frac{\p \tbeta}{\p r}.
\ee
Under the condition that there is no {\it dark matter} in the system under consideration, the above equation reduces to the one found in Ref. \cite{abd20}.

The exact form of the tetrad for the fiducial comoving observer, derived in Ref. \cite{fel03} and the values of the magnetic components
in the tetrad in question, enable us to find the exact form of $\cL(\la_{\hat 0})$.
Namely it is provided by the relation $\cL(\la_{\hat 0}) = e^\psi f(r)$,
where we have denoted
\be 
e^\psi = \Big( f(r) - \Omega^2r^2 \Big)^{-\frac{1}{2}}.
\label{ll}
\ee
The angular velocity of the orbital motion measured by the observer located at infinity, is given by 
$\Omega= \frac{d \phi}{dt} = \frac{f(r)}{r^2} \frac{1}{\cE}$.

Having in mind the adequate components of the fiducial observer tetrad \cite{fel03}
one can find that the non-zero components of the {\it visible} and {\it dark sector} magnetic fields, seen by an observer in a comoving plane at $\theta= \pi/2$,
are given by
\be
B_{\hat{\theta}}^{(\tF)}= B_0^{(\tF)} f(r) e^\psi, \qquad B_{\hat{\theta}}^{(\tB)}= B_0^{(\tB)} f(r) e^\psi.
\ee

Finally, using the relations for  $\cL(\la_{\hat 0})$ and (\ref{ll}),
we arrive at the exact form of the magnetic coupling $\tilde{\beta}$, which envisages that a magnetized particle possessing energy $\cal E$
and angular momentum $l$, can move on a circular orbit around black hole in question at the adequate distance $r$
\be 
\label{beta}
\tilde{\beta} = \left( \frac{1}{f(r)} - \frac{l^2}{\mathcal{E}^2 r^2  }   \right)^{1/2} \left( 1 + \frac{l^2}{r^2} - \frac{\mathcal{E}^2}{f(r)}  \right).
\ee
From equation (\ref{bbd}) it can be deduced that {\it dark sector} influences the particle energy and momentum $l$, by means of {\it dark sector magnetic} 
field, as well as, $\alpha$ coupling constant.
Namely, using the relation (\ref{bbd}), one can find
\be 
\label{l-r}
\frac{l^2}{r^2}= \cL(\la_{\hat 0}) \Big( \beta(F) + \beta(B) + \frac{\alpha}{2}  (\beta(F,B) + \beta(B,F)) \Big) + \frac{\mathcal{E}^2}{f(r)} - 1.
\ee
Further,
on substituting the equation (\ref{l-r}) into the relation (\ref{beta}), we finally arrive at the following:
\ben \nonumber
\label{beta1}
\tilde{\beta} &=& \frac{\sqrt{f(r)}}{\mathcal{E}} \Big( \beta(F) + \beta(B) + \frac{\alpha}{2}  (\beta(F,B) + \beta(B,F)) \Big) \\
&\times& \sqrt{1 - \Big( \beta(F) + \beta(B) + \frac{\alpha}{2}  (\beta(F,B) + \beta(B,F)) \Big) \sqrt{f(r)}}.
\een
In the above expression, we have used the fact that $\cL(\la_{\hat 0}) = \sqrt{f(r)}$ for a comoving observer. 

\begin{figure}[H]   
\centerline{ \includegraphics[width=12cm]{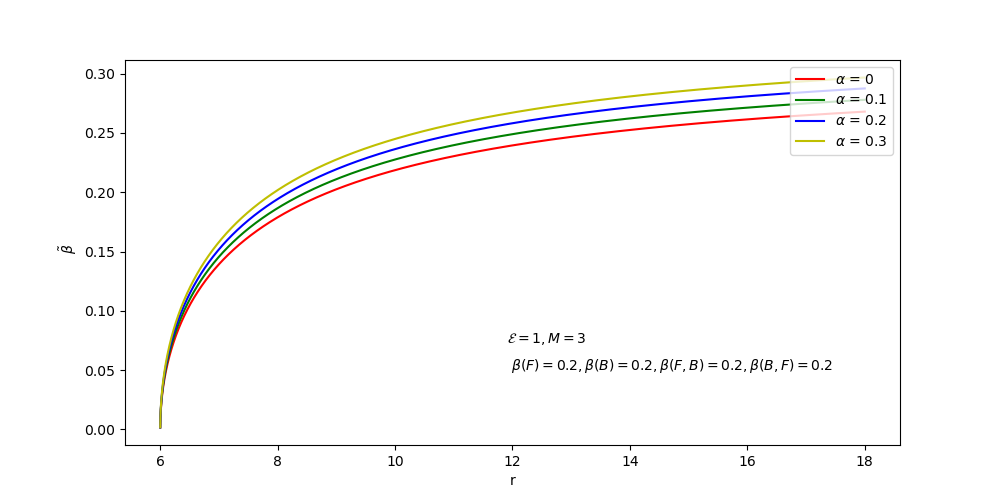} }
\caption{(color on line) Dependence of $\tilde{\beta}$ on $r$ for different values of $\alpha$-coupling constant. }
\label{Fig:0}
\end{figure}

The radial dependence of $\tilde{\beta}(r)$ envisages the fact that with the growth of the distance from black hole event horizon, one observes 
enlargement of $\tilde{\beta}(r)$ value and the higher value of $\alpha$ we analyze the bigger $\tilde{\beta}(r)$ one gets. The bunch of curves corresponding to the different 
values of $\alpha$ separates for large $r$. We remark that because of the fact that the coupling constant between two sectors is very small, in what follows
we shall be interested in qualitative behaviors of $\alpha$-depending phenomena, taking into account values which can envisage the $\alpha$ way of acting.


\section{Collisions of particles in the nearby of black hole}

In this section, we discuss the collision of two particles in the vicinity of a 
weakly magnetized by {\it visible} and {\it hidden sector} fields, static spherically symmetric black hole, looking for the effect of
{\it dark photon} on the process in question.
The main idea is to solve the Hamilton-Jacobi equation (\ref{HJ-eq}) using the substitution (\ref{sol-HJ}), and obtain the four-velocity components of a particle
moving in the nearby of black hole. To begin with
we shall consider action  (\ref{sol-HJ}) written in the following form:

\ben
S&=& -E~t + L~ \phi \\ \nonumber
&+& \int dr \sqrt{ g(r) \left[  \frac{E^2}{f(r)}  - \frac{1}{r^2} \left( L - \tilde{e}_{A} \tilde{A}_{\phi} - \tilde{e}_{B} \tilde{B}_{\phi} \right)^2 - m^2 + 
m \Big(\tD^{\mu \nu}_{(\tF)} ~\tF_{\mu \nu} + \tD^{\mu \nu}_{(\tB)} ~ \tB_{\mu \nu} \Big) \right] } .
\een
The trajectory of the particle is determined by $\frac{\partial{S}}{\partial{L}} = constant$, 
while the time dependence is determined by $\frac{\partial{S}}{\partial{E}} = constant$. 
Because of the fact that the metric is independent of time, the covariant component of the four-velocity $u_0 = -\frac{E}{m}$ is conserved. This results in 
relation provided by
\be
 \frac{dt}{d\tau} = \frac{\cE}{f(r)}.
\ee
The other components of the four-velocity in the plane $\theta= \pi/2$ imply
\ben
\Big( \frac{dr}{d \tau} \Big)^2 &=&
f(r) \Bigg[ \frac{\cE^2}{f(r)} - \frac{1}{r^2} \Big( l - \frac{\te_A \tA_\phi }{m} - \frac{\te_B \tB_\phi }{m} \Big)^2 -1 
+ \frac{D^{\mu \nu}_{(\tF)} \tF_{\mu \nu} + D^{\mu \nu}_{(\tB)} \tB_{\mu \nu} }{m} \Bigg], \\
\frac{d \phi}{d \tau} &=& \frac{1}{r^2} \Big( l - \frac{\te_A \tA_\phi }{m} - \frac{\te_B \tB_\phi }{m} \Big).
\een
On the other hand, the center of mass energy of the particles yield
\be
\cE_{CM}^2 = 1 - g_{\mu \nu} u^{\mu}_1~u^{\nu}_2.
\ee
Its explicit form is provided by
\ben
\cE_{CM}^2 &=&
1 + \frac{\cE_1~\cE_2}{f(r)} 
- \frac{1}{r^2} \Big(l_1 - \frac{\te_A \tA_\phi }{m} - \frac{\te_B \tB_\phi }{m} \Big)\Big(l_2 - \frac{\te_A \tA_\phi }{m} - \frac{\te_B \tB_\phi }{m} \Big) \\ \nonumber
&-& \frac{1}{f(r)}
\Bigg[ \cE^2_1 - \frac{f(r)}{r^2} \Big( l_1 - \frac{\te_A \tA_\phi }{m} - \frac{\te_B \tB_\phi }{m} \Big)^2 -1 
+ \frac{f(r) \Big(D^{\mu \nu}_{(\tF)} \tF_{\mu \nu} + D^{\mu \nu}_{(\tB)} \tB_{\mu \nu}\Big)_{(1)} }{m} \Bigg]^{\frac{1}{2}} \\ \nonumber
&\times&
\Bigg[ \cE^2_2 - \frac{f(r)}{r^2} \Big( l_2 - \frac{\te_A \tA_\phi }{m} - \frac{\te_B \tB_\phi }{m} \Big)^2 -1 
+ \frac{f(r) \Big(D^{\mu \nu}_{(\tF)} \tF_{\mu \nu} + D^{\mu \nu}_{(\tB)} \tB_{\mu \nu}\Big)_{(2)} }{m} \Bigg]^{\frac{1}{2}}.
\een
In the subsequent sections we shall pay attention to the collision scenarios which can be valid from the point of view of the possible future
detection of {\it dark sector}, envisaging the influence of $\alpha$-coupling constant on the process in question.

\subsection{Two Magnetized particles}
We shall commence with the magnetized particles moving in equatorial plane around weakly magnetized Schwarzschild black hole. The four-velocity components imply the following relations:
\ben
 \frac{dt}{d\tau} &=& \frac{\cE}{f(r)}, \\
 \Big( \frac{dr}{d \tau} \Big)^2 &=&
\cE^2 - f(r) \Big[ \frac{l^2}{r^2} + 1 - \sqrt{f(r)} ~\tbeta(F, B, \alpha) \Big],\\
 \frac{d \phi}{d \tau} &=& \frac{l}{r^2}.
 \een
Consequently the center mass energy is given by
\ben
\cE_{CM}^2 &=&
1 + \frac{\cE_1~\cE_2}{f(r)}  - \frac{1}{f(r)} \Big[ \cE_1^2 - f(r) \Big(
\frac{l_1^2}{r^2} +1 - \sqrt{f(r)} \tbeta_1(F, B, \alpha) \Big)\Big]^{\frac{1}{2}} \\ \nonumber
&\times&
\Bigg[ \cE^2_2 - f(r) \Big(
\frac{l_2^2}{r^2} +1 - \sqrt{f(r)} \tbeta_2(F, B, \alpha)\Big)  \Big]^{\frac{1}{2}} - \frac{l_1~l_2}{r^2}.
\een
One notices that the parameter $l$ can be positive or negative. It means that for $l > 0$, the Lorentz force acting on a charged particle
is repulsive, while for $l<0$ is attractive.  The repulsive Lorentz force is directed outward considered black hole, for the attractive
case its direction is toward black hole.

\begin{figure}[H]   
\centerline{ \includegraphics[width=12cm]{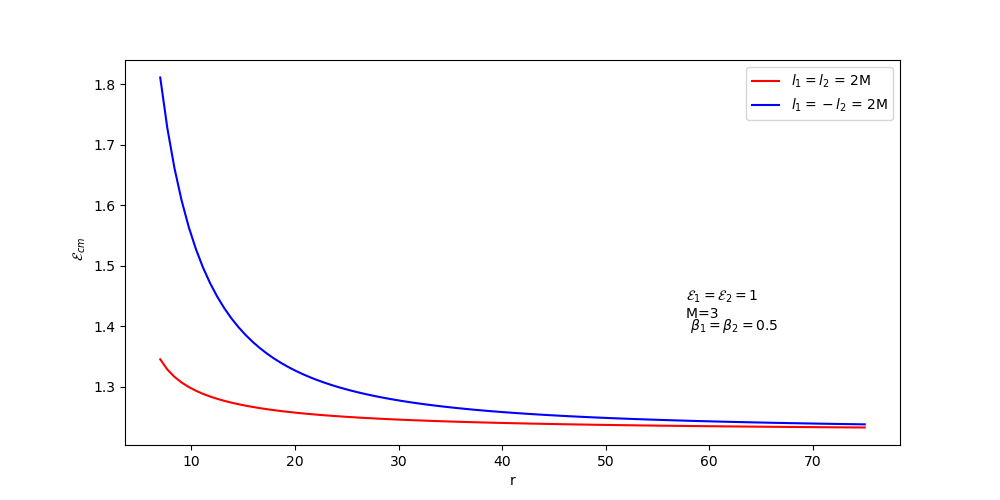} }
\caption{(color on line) $\cE_{CM}$ of two magnetized particles as a function of $r$. $M=3$ is the mass of the black hole, $l_1= l_2=2 M$ and
$l_1= - l_2=2 M$, where $l_{1,2}$
 are the angular momenta per unit mass of the first and second particles.}
\label{Fig:1}
\end{figure}

In  Fig. \ref{Fig:1} 
we present the $\cE_{CM}$ dependence on the distance $r$, for two magnetized particle. In the case of $l_1 = l_2 = 2M$ the energy grows more rapidly
when one moves closer to the black hole event horizon, comparing to the case $l_1 = - l_2 = 2M$.

\subsection{Magnetized and charged particles}
Now we shall consider collisions of the magnetized and charged particles in the background of the aforementioned black hole.
The velocity components yield
\ben
 \frac{dt}{d\tau} &=& \frac{\cE}{f(r)}, \\
 \Big( \frac{dr}{d \tau} \Big)^2 &=&
f(r) \Bigg[ \frac{\cE^2}{f(r)} - \frac{1}{r^2} \Big( l - e r^2 ( w^{(F)} + \frac{\alpha}{2} w^{(B)}) \Big)^2 -1 \Bigg],\\
 \frac{d \phi}{d \tau} &=& \frac{1}{r^2} \Big( l - e r^2 ( w^{(F)} + \frac{\alpha}{2} w^{(B)}) \Big),
 \een
where we have denoted
\be
w^{(F)} = \frac{B_0^{(F)}}{2 m}, \qquad w^{(B)} = \frac{B_0^{(B)}}{2 m}.
\label{wfwb}
\ee
In the case under consideration the center of mass energy is provided by
\ben
\cE_{CM}^2 &=&
1 + \frac{\cE_1~\cE_2}{f(r)}  - \frac{1}{f(r)} \Big[ 
\cE_1^2 - \frac{f(r)}{r^2} \Bigg( l_1 - e r^2 \Big( w^{(F)} + \frac{\alpha}{2} w^{(B)}\Big) \Bigg)^2 -f(r) \Big]^{\frac{1}{2}} \\ \nonumber
&\times&
\Bigg[ \cE^2_2 - f(r) \Big(
\frac{l_2^2}{r^2} +1 - \sqrt{f(r)} \tbeta_2(F, B, \alpha) \Big)\Big]^{\frac{1}{2}} 
- \frac{1}{r^2} \Bigg( l_1 - e r^2 \Big( w^{(F)} + \frac{\alpha}{2} w^{(B)}\Big) \Bigg) l_2.
\een

\begin{figure}[H]   
\centerline{ \includegraphics[width=12cm]{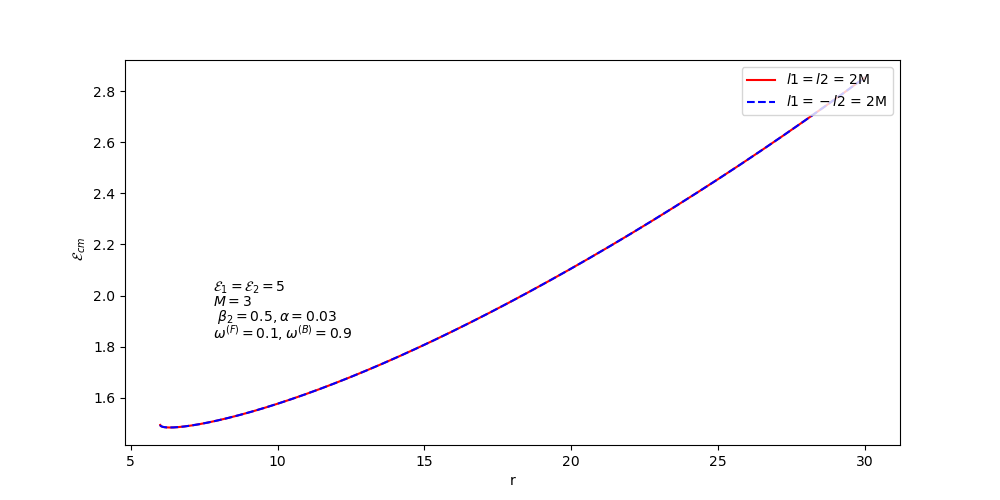} }
\caption{(color on line) $\cE_{CM}$ as a function of $r$, for two particles, one magnetized and one charged, where $\cE_1 = 5$ and $\cE_1 = 5$ are the energies per unit mass of particles 1 and 2, respectively.  }
\label{Fig:3}
\end{figure}

\begin{figure}[H]   
\centerline{ \includegraphics[width=12cm]{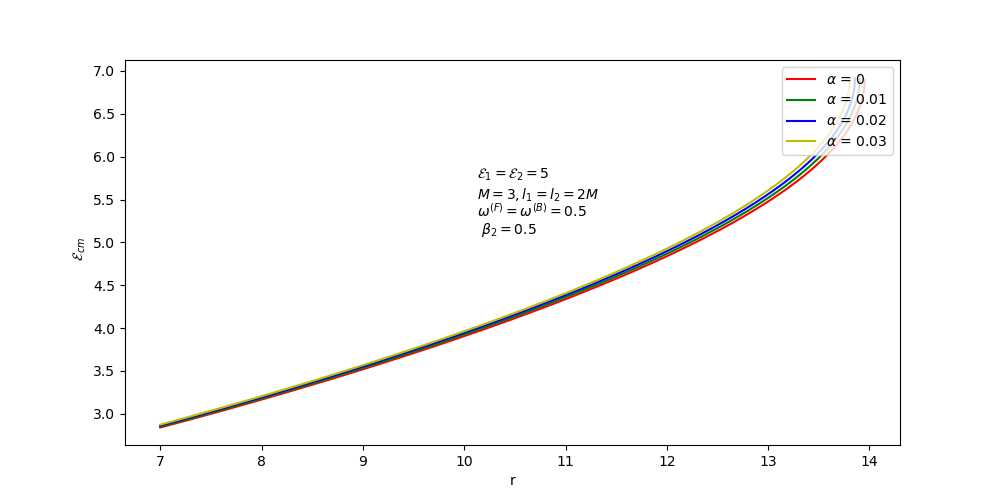} }
\caption{(color on line) The dependence of $\cE_{CM}$ of two particles, one magnetized and one charged,  on $ r$ and for various values of $\alpha$-coupling constant. The rest of the parameters are the same as in Fig. \ref{Fig:3} .} 
\label{Fig:4}
\end{figure}

In Fig. \ref{Fig:3} we have the dependence of $\cE_{CM}$ on $r$, for charged and magnetized particles, having the same energy and 
charge due to Maxwell and {\it dark photon } fields, the magnetic parameter is equal to $0.5$. 
One obtains that for the large distance from black hole event horizon, the energy grows. 
For the case $l_1 = l_2 $ and $l_1 = - l_2$, figures overlap. 

In Fig. \ref{Fig:4}, the plot was done for various values of $\alpha$-coupling constant.  We have that the larger $\alpha$ one elaborates the bigger value of 
$\cE_{CM}$ we get. 

\begin{figure}[H]   
\centerline{ \includegraphics[width=12cm]{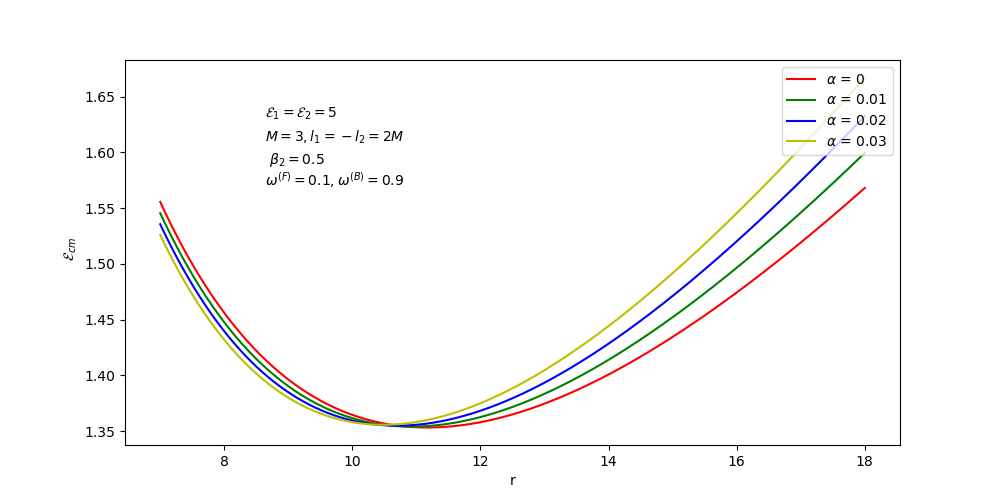} }
\caption{(color on line)  $\cE_{CM}$ on $r$-dependence of two particles (one magnetized one charged) for high concentration of {\it dark matter} and various values of $\alpha$-coupling constant.}
\label{Fig:5}
\end{figure}

Fig. \ref{Fig:5} is devoted to the same kind of particles ($l_1 = -l_2$) but under the assumption that one has greater concentration of {\it dark matter}, i.e., $w^{(B)}> w^{(F)}$.
In the elaborated case we obtain bunch of curves for different values of $\alpha$-coupling constant cutting in one point  $r_c$ (the so-called isosbestic point, see , e.g. \cite{kic20}).
For the distance $r > r_c$ one achieves the result that 
the larger $\cE_{CM}$ is achieved for bigger values of coupling constant. On the contrary for $r< r_c$ the situation is opposite.

\begin{figure}[H]   
\centerline{ \includegraphics[width=12cm]{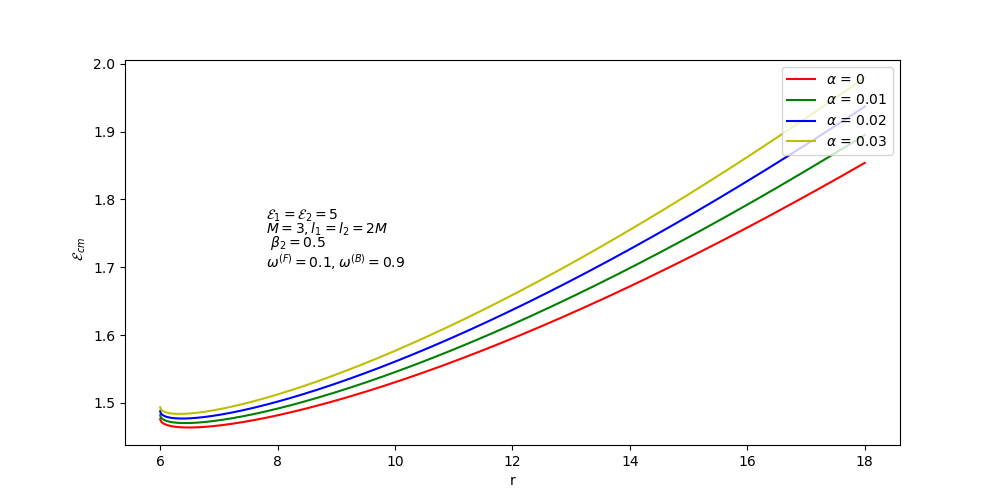} }
\caption{(color on line) $\cE_{CM}$ on $r$-dependence of two particles, one magnetized and one charged,  for high value of {\it dark photon} field. The repulsive Lorentz force case
with different values of $\alpha$-coupling constant.}
\label{Fig:6}
\end{figure}

In Fig. \ref{Fig:6}, for the case $l_1=l_2$, we have that the larger $\alpha$ we study, the bigger energy one obtains.

On the other hand, for the high values of the Maxwell fields, we have no difference among the curves connected with different values of $\alpha$-coupling constant.
The energy grows with the distance for $l_1 = - l_2$.

\begin{figure}[H]   
\centerline{ \includegraphics[width=12cm]{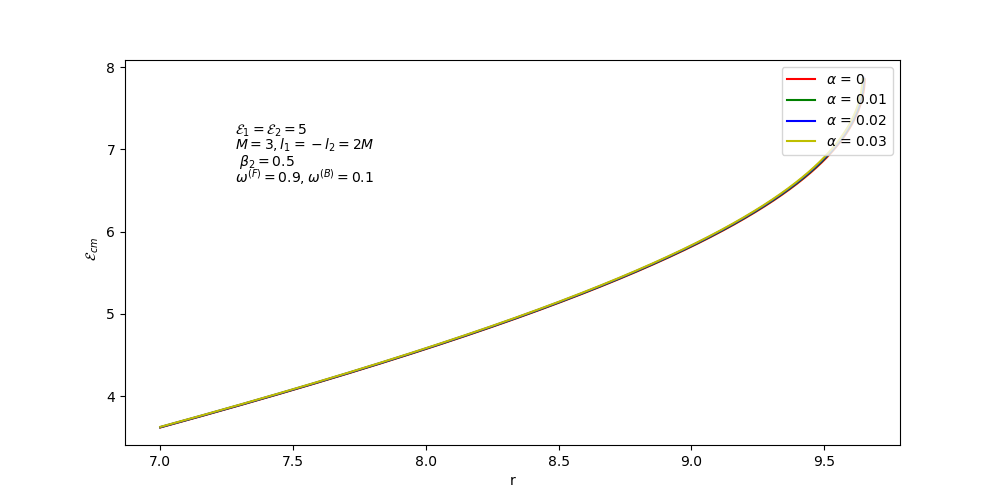} }
\caption{(color on line) Two particles, one magnetized and one charged, 
$\cE_{CM}$ dependence on the distance, for high value of Maxwell field around black hole.}
\label{Fig:7}
\end{figure}

In the case of
small values of gauge fields and the same parameters as in Fig. \ref{Fig:7}, one has that the tendency reverses, i.e., the energy $\cE_{CM}$ grows with the
diminishing of the distance to the black hole event horizon.
\begin{figure}[H]   
\centerline{ \includegraphics[width=12cm]{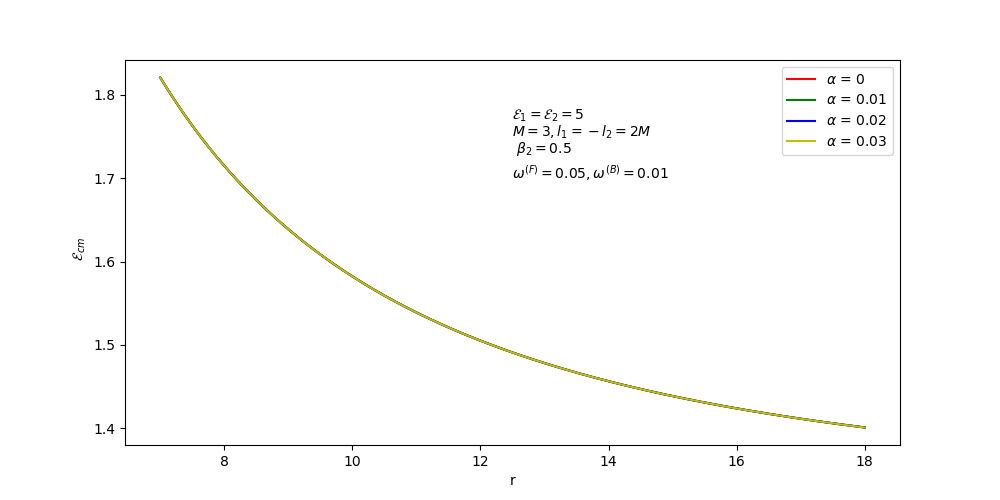} }
\caption{(color on line) The same functional dependence as in Fig. 7, but for small values of Maxwell and {\it dark photon} fields.}
\label{Fig:8}
\end{figure}

\subsection{Neutral and magnetized particles}
It happens that
for the neutral and magnetized particle collision the description will be qualitatively the same as presented in Ref. \cite{abd20}, i.e., the same account as for the ordinary Schwarzschild black hole spacetime,
but quantitatively one gets the different values of $\cE$ and $l$, due to the fact that we take also into account magnetic field stemming from {\it dark matter sector}.

\section{Charged particle motion}
In this section we shall elaborate the problem of a charged particle motion in the spacetime of a weakly magnetized black hole.
As in the previous sections, a particle will be charged under two $U(1)$-gauge sectors, {\it visible} and {\it dark} one. 
One elaborates several cases of magnetic field, homogeneous at spatial infinity, dipolar, parabolic and magnetic 
field stemming from current loops (Maxwell field and {\it dark photon} one)
surrounding Schwarzschild black hole.

\subsection{Constant homogeneous magnetic field.}
To commence with, 
we suppose that there exist magnetic fields stemming 
from those sectors in the nearby of black hole, being axisymmetric and homogenous at spatial infinity.

Having in mind the generalized four-momentum of the particle (\ref{mom}), the conserved quantities bounded with the symmetry of the studied problem,
respectively
specific energy and azimuthal angular momentum, will be 
provided by
\ben \label{def1}
\cE &=& - \frac{1}{m} \xi^\mu_{(t)} {\tilde p}_\mu = {\dot t}~f(r),\\ \label{def2}
l_z &=& \frac{1}{m} \xi^\mu_{(\phi)} {\tilde p}_\mu = r^2 \sin^2 \theta \Big( {\dot \phi} + \te_A \tB_0^{(\tF)} + {\te_B} \tB_0^{(\tB)} \Big),
\een
where by the dot we denote derivation with respect to time,  $d/ d\tau$ and set
\be
\tB_0^{(\tF)} = \frac{e}{2m} B_0^{(\tF)}, \qquad \tB_0^{(\tB)} = \frac{e_d}{2m} B_0^{(\tB)},
\label{stat1}
\ee
and
\be
B^{(\tF)}_0 = \frac{\sqrt{2 - \alpha}}{2} \Big( B^{(F)}_0 - B^{(B)}_0 \Big), \qquad
B^{(\tB)}_0 = \frac{\sqrt{2 + \alpha}}{2} \Big( B^{(F)}_0 + B^{(B)}_0 \Big).
\ee
The normalization condition of the four-velocity, $u_\alpha u^\alpha = -1$ leads to the relation
\be
\cE^2 = {\dot r}^2 + r^2 f(r) {\dot \theta}^2 + U_{eff},
\label{ee}
\ee
where we have defined the effective potential in the following form:
\be
U_{eff} = f(r) \Big[ 1 + r^2 \sin^2 \theta \Big( \frac{l_z}{r^2 \sin^2 \theta} - \te_A \tB_0^{(\tF)}  -  \te_B \tB_0^{(\tB)} \Big)^2 \Big].
\label{uffo}
\ee
For the case when $e_d =0$, the effective potential in question yields
\be
U_{eff} = f(r) \Bigg[ 1 + r^2 \sin^2 \theta \Bigg( \frac{l_z}{r^2 \sin^2 \theta} - {e}\Big( w^{(F)} + \frac{\alpha}{2} w^{(B)} \Big) \Bigg)^2
 \Bigg].
\label{uff}
\ee
If one restricts his considerations to the case of an equatorial plane where $\theta = \pi/2$ (see Fig. 10), we observe that the bigger value of $\alpha$-coupling
constant one takes, the greater $U_{eff}$ one gets. For the case $B^{(F)}_0 = B^{(B)}_0$ we have clear distinction among the curves which grows with the distance from
black hole event horizon.

\begin{figure}[H]   
\centerline{ \includegraphics[width=12cm]{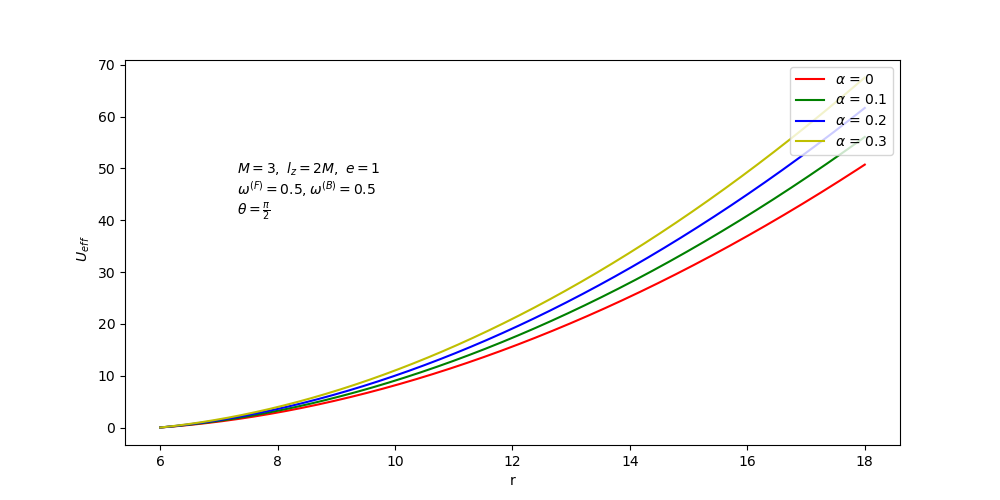} }
\caption{(color on line) The dependence of $U_{eff}$ on $r$ for different values of $\alpha$-coupling constant. }
\label{Fig:11}
\end{figure}

The relation (\ref{ee}) is constraint, i.e., it is satisfied at the initial starting time and with the passage of time is always fulfilled, guiding the dynamics of 
$r(\tau)$ and $\theta( \tau)$.

The $r$ and $\theta$ components of the equation of motion (\ref{eqmot}) are given, respectively by
\ben \nonumber
{ \ddot r} &+& \Big( r {\dot \theta}^2 
+ \frac{l_z^2}{r^3 \sin^2 \theta} \Big) \Big(  - f(r) + \frac{r f'(r)}{2} \Big)
+ \Big(  \te_A \tB_0^{(\tF)}  + \te_B  \tB_0^{(\tB)} \Big)^2 ~\Big(  f(r) + \frac{r f'(r)}{2} \Big) r \sin^2 \theta \\ 
&+& f'(r)
\Big[ \frac{1}{2} -  l_z \Big( \te_A \tB_0^{(\tF)}  +  \te_B \tB_0^{(\tB)} \Big) \Big] = 0, \\
{\ddot \theta} &=& - \frac{2}{r} ~{\dot r} {\dot \theta} + \frac{l_z^2 \cos \theta}{r^4 \sin^3 \theta} - \Big( \te_A \tB_0^{(\tF)}  +  \te_B \tB_0^{(\tB)} \Big)^2 \sin \theta \cos \theta.
\een

Restricting our consideration to the case $e_d=0$, one obtains the following equations:
\ben \nonumber
{ \ddot r} &+& \Big( r {\dot \theta}^2 
+ \frac{l_z^2}{r^3 \sin^2 \theta} \Big) \Big(  - f(r) + \frac{r f'(r)}{2} \Big)
+ {e^2}\Big( w^{(F)}  + \frac{\alpha}{2} w^{(B)} \Big)^2 ~\Big(  f(r) + \frac{r f'(r)}{2} \Big) r \sin^2 \theta \\ 
&+& f'(r)
\Big[ \frac{1}{2} - {e~ l_z}\Big( w^{(F)}  +  \frac{\alpha}{2} w^{(B)} \Big) \Big] = 0, \\
{\ddot \theta} &=& - \frac{2}{r} ~{\dot r} {\dot \theta} + \frac{l_z^2 \cos \theta}{r^4 \sin^3 \theta} - 
{e^2}
\Big( w^{(F)}  +  \frac{\alpha}{2} w^{(B)} \Big)^2 \sin \theta \cos \theta.
\een
From the above it can be clearly seen that {\it dark sector} modifies the geodesic equations by its  own magnetic field and $\alpha$-coupling constant.

The position of the innermost stable circular orbit (ISCO) is determined by $\partial_{r} U_{eff} = \partial^{2}_{r} U_{eff} =0$. 
Consequently, 
following the approach provided in Ref.\cite{fro10}, one gets
the following relations:
\ben \nonumber
\label{r-pm}
e\Big( w^{(F)}  +  \frac{\alpha}{2} w^{(B)} \Big) &=& \frac{1}{\sqrt{2} \sin \theta} \\
&\times& \frac{\sqrt{2M(6M-r_{\pm})}}{r_{\pm} \Big[ 4 r^{2}_{\pm} - 18 M r_{\pm} + 12 M^2 \pm 2M \sqrt{\big( 3 r_{\pm} - 2M  \big)  \big( 6M - r_{\pm}  \big) } \big]^{1/2} },
\een
and
\ben 
l_{\pm} &=& \pm \frac{1}{\sqrt{2}}\frac{\sqrt{2M} (\sin \theta) r_{\pm} }{r_{\pm} \Big[ 4 r^{2}_{\pm} - 18 M r_{\pm} + 12 M^2 \pm 2M \sqrt{\big( 3 r_{\pm} - 2M  \big)  \big( 6M - r_{\pm}  \big) } \big]^{1/2}},
\een
where $ r_{\pm} \equiv \sqrt{r^2} $. On the other hand, the Lorentz $\gamma$ factor of a charged particle in a circular orbit, as defined in Ref. \cite{fro12}, yields
\ben 
\gamma_{\pm} &=& \frac{2 \big( r_{\pm} - 2M \big)}{\Big[ 4 r^{2}_{\pm} - 18 M r_{\pm} + 12 M^2 \pm 2M \sqrt{\big( 3 r_{\pm} - 2M  \big)  \big( 6M - r_{\pm}  \big) } \big]^{1/2}}.
\een
It can be seen that
the above Lorentz factor is exactly the same as in the case of a charged particle moving on a ISCO under Maxwell's weak magnetic field. The absence of any explicit dependence of 
{\it dark photon} coupling in the dynamics of a charged particle excludes the possibility to observe any deviations using an analytical expression.  

However, 
in order to envisage the dependency of ISCO on $\alpha$-coupling constant, let us rewrite equation  (\ref{r-pm}) in terms of two newly defined functions $ Q^{+} $  and  $Q^{-}$
provided by 
\ben \nonumber
\label{Q-plus}
Q^{+}(r_{\pm})  &\equiv&  4 r_{\pm}^4 -18M r_{\pm}^3 + 12 M r_{\pm}^2 + 2M r_{\pm}^2 \sqrt{(3r_{\pm} - 2M)(6M-r_{\pm})} \\  
&-&   \frac{M(6M-r_{\pm})}{ e^2\Big( w^{(F)}  +  \frac{\alpha}{2} w^{(B)} \Big)^2 \sin^2 \theta},
\een
and
\ben \nonumber
\label{Q-minus}
Q^{-}(r_{\pm})  &\equiv&  4 r_{\pm}^4 -18M r_{\pm}^3 + 12 M r_{\pm}^2 - 2M r_{\pm}^2 \sqrt{(3r_{\pm} - 2M)(6M-r_{\pm})} \\  
&-&   \frac{M(6M-r_{\pm})}{ e^2\Big( w^{(F)}  +  \frac{\alpha}{2} w^{(B)} \Big)^2 \sin^2 \theta}.
\een
It can be seen that $Q^{+}$ and $Q^{-}$ differ by the the choice of $+$ or $-$ sign in the denominator of the relation (\ref{r-pm}). 
In order to obtain the dependency of the ISCO on $\alpha$, we plot $Q^{+}$ and $Q^{-}$ as a function of $r_{\pm}$ and get the points where a horizontal line passing through the origin intersects these curves. 

\begin{figure}[H]   
\centerline{ \includegraphics[width=12cm]{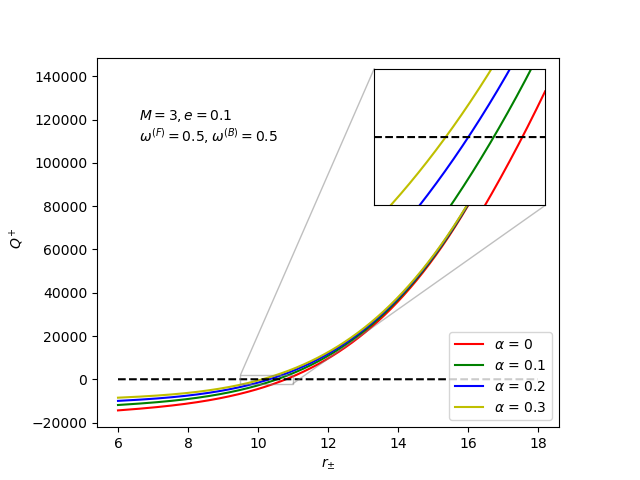} }
\caption{(color on line) The plot showing the dependence of $Q^{+}$ on $r_{\pm}$. The points on the curve intersected by the horizontal dashed line are the roots $Q^{+}$. }
\label{Fig:9}
\end{figure}

\begin{figure}[H]   
\centerline{ \includegraphics[width=12cm]{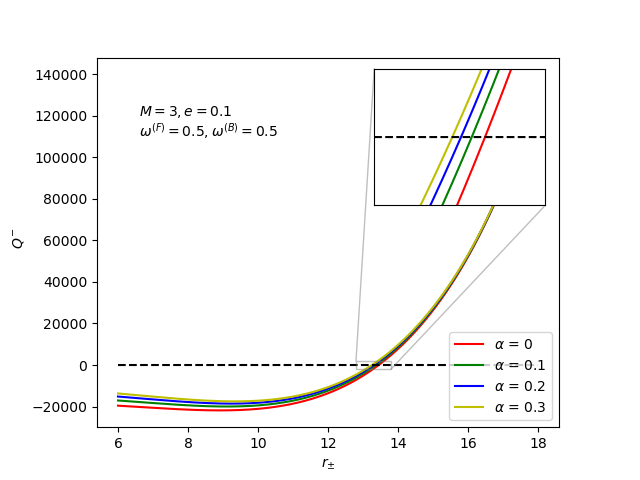} }
\caption{(color on line) The plot showing the dependence of $Q^{-}$ on $r_{\pm}$. The points on the curve intersected by the horizontal dashed line are the roots $Q^{-}$. }
\label{Fig:10}
\end{figure}

The inspection of Fig. \ref{Fig:9} and \ref{Fig:10} reveals
that ISCO are influenced by the presence of {\it dark photons}. Namely the radius of ISCO decreases as the coupling between 
{\it dark photon} and Maxwell fields increases.

The ISCO plays an important role in black hole accretion disks since it marks the inner edge of the accretion disk and in principle 
the difference in the radius of ISCO caused by the {\it dark sector} may be observed. But because of the small value of $\alpha$-coupling constant
(the analysis of Planck CMB and unWISE galaxy survey \cite{mcc24}
reveal constraints on {\it dark photon} parameter $\alpha \le 4.5 \times 10^{-8}$ (95 percent of confidence level),
as well as, the range of its mass $ 10^{-13}  eV \le m_{dark ~ph} \le 10^{-11} eV$) this task is beyond the contemporary observations and
will constitute a difficult challenge to 
future generations of Event Horizon Telescope collaboration observations.

\subsection{Dipolar and parabolic configurations}
In this subsection we shall elaborate the case when the magnetic field which surrounds the black hole in question is dipolar/ parabolic.
The components of the adequate four-potentials are given respectively by the relations \cite{wal74}-\cite{qi23}
 \be 
 \tA^\mu = \frac{ B_0^{(\tF)}}{2} G_k(r)~H_k(\theta)~\xi^\mu_{(\phi)}, \qquad  \tB^\mu = \frac{B_0^{(\tB)}}{2}G_k(r)~H_k (\theta)~\xi^\mu_{(\phi)},
\ee
where $B_0^{{(\tF) (\tB)}}$ are given by (\ref{bbb}), while for $k = d, ~p$ for dipolar and parabolic configurations
\ben
G_d(r) &=& r^2 \Big[ \ln f(r) + \frac{2M }{r} \Big( 1 + \frac{M}{r} \Big) \Big], \qquad H_d(r) = \sin^2 \theta,\\
G_p(r) &=& r^s, \qquad H_p(\theta) = 1 - \mid \cos \theta \mid.
\een
$s= 3/4$ corresponds to the parabolic configuration of magnetic fields..

For $e_d =0$, using the definitions (\ref{def1})-(\ref{def2}) and the normalization condition for four-velocity, one arrives at the following form of the effective potential:
\be
U_{eff} = f(r) \Bigg[ 1 + r^2 \sin^2 \theta \Bigg( \frac{l_z}{r^2 \sin^2 \theta} - {e}\Big( {\tilde w}^{(F)} + \frac{\alpha}{2} {\tilde w}^{(B)} \Big) \Bigg)^2
 \Bigg],
\ee
where we set
\be
{\tilde w}^{(F)} = \frac{B_0^{(F)} G_k(r) H_k(\theta)}{2 m}, \qquad {\tilde w}^{(B)} = \frac{B_0^{(B)}G_k(r) H_k(\theta)}{2 m}.
\ee
The specific forms of effective potential for the dipolar and parabolic configurations for the equatorial plane, $\theta = \frac{\pi}{2}$, are given by
\be
U^{(d)}_{eff} = f(r) \Bigg[ 1 + r^2  \Bigg( \frac{l_z}{r^2} - {e} \frac{r^2}{2m} \left( \ln f(r) +\frac{2M}{r} \left( 1 + \frac{M}{r}\right)    \right)    \Big( B_0^{(F)} + \frac{\alpha}{2} B_0^{(B)} \Big) \Bigg)^2
 \Bigg],
\ee

\be
U^{(p)}_{eff} = f(r) \Bigg[ 1 + r^2  \Bigg( \frac{l_z}{r^2} - {e} \frac{r^{\frac{3}{4}}}{2m}    \Big( B_0^{(F)} + \frac{\alpha}{2} B_0^{(B)} \Big) \Bigg)^2
 \Bigg],
\ee

\begin{figure}[H]   
\centerline{ \includegraphics[width=12cm]{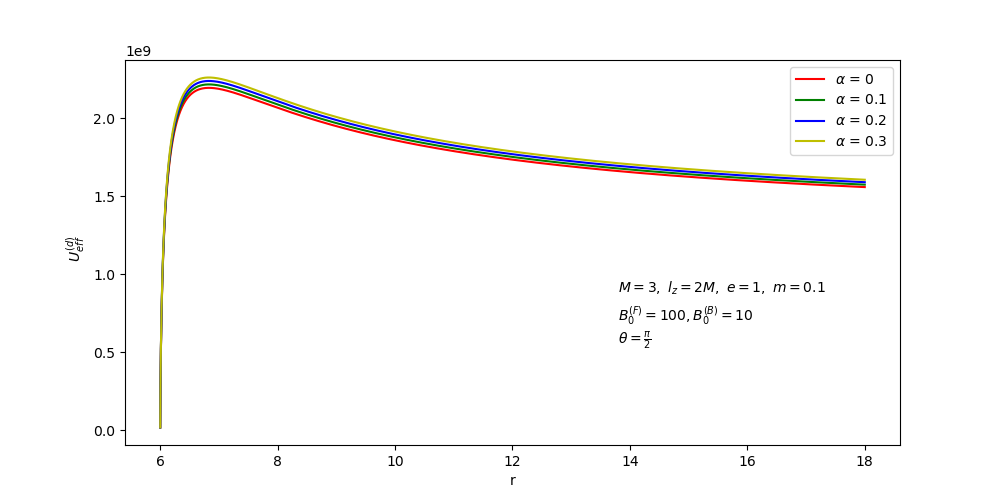} }
\caption{(color on line) The dependence of $U^{(d)}_{eff}$ on $r$ for different values of $\alpha$ in the case of dipolar configuration of magnetic fields, for the dominance
of Maxwell field, 
$B^{(F)}_0 > B^{(B)}_0$. }
\label{Fig:12}
\end{figure}

\begin{figure}[H]   
\centerline{ \includegraphics[width=12cm]{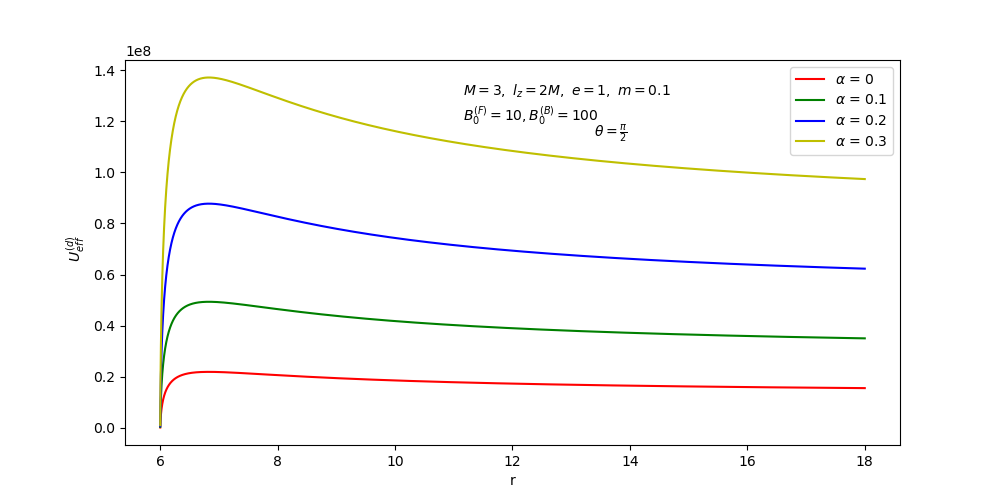} }
\caption{(color on line) The dependence of $U^{(d)}_{eff}$ on $r$ for different values of $\alpha$ in the case of dipolar configuration of magnetic fields, for 
the domination of {\it dark photon} field, i.e., $B^{(F)}_0 < B^{(B)}_0$. }
\label{Fig:13}
\end{figure}

\begin{figure}[H]   
\centerline{ \includegraphics[width=12cm]{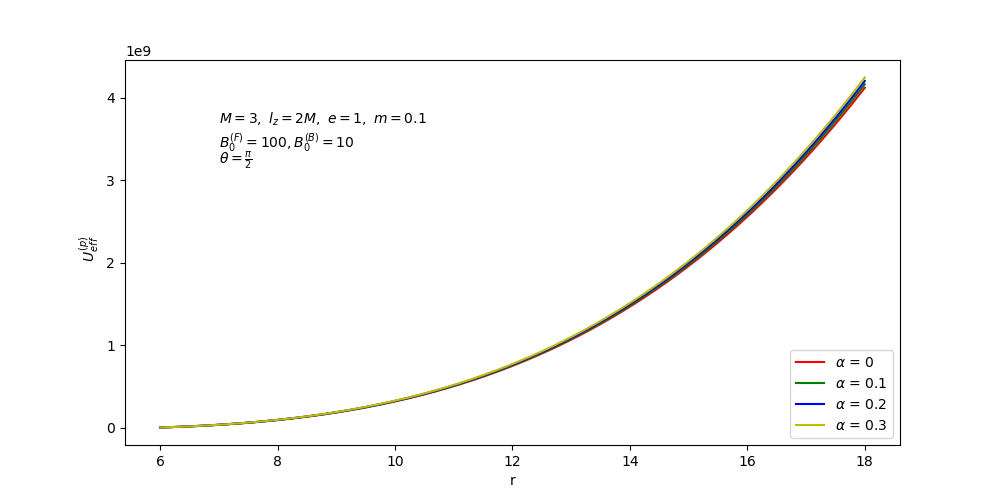} }
\caption{(color on line) The effective potential for a parabolic magnetic field configuration  $U^{(p)}_{eff}$ as a function of $r$, 
for different values of $\alpha$, when $B^{(F)}_0 > B^{(B)}_0$. }
\label{Fig:14}
\end{figure}

\begin{figure}[H]   
\centerline{ \includegraphics[width=12cm]{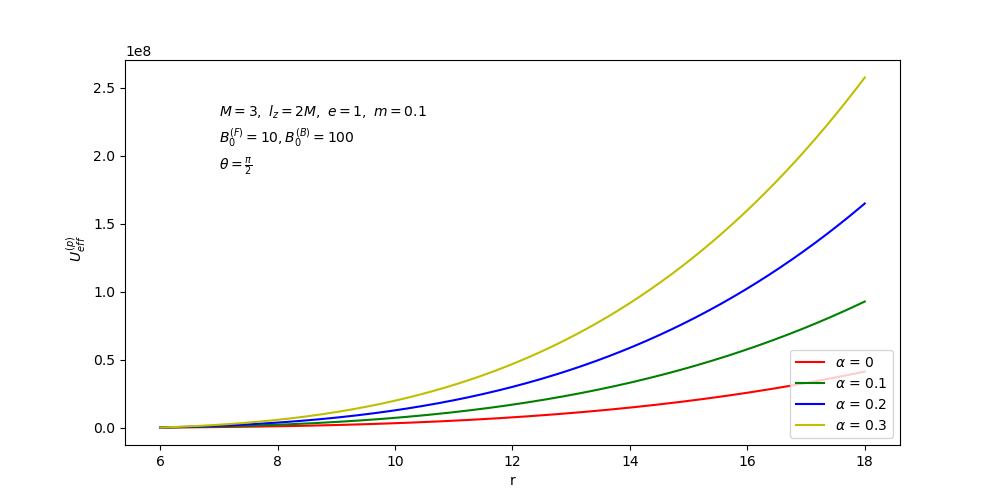} }
\caption{(color on line) The effective potential for a parabolic magnetic field configuration  $U^{(p)}_{eff}$ as a function of $r$, 
for different values of $\alpha$, when $B^{(F)}_0 < B^{(B)}_0$. }
\label{Fig:15}
\end{figure}

Figs. \ref{Fig:12}-\ref{Fig:15} envisage the influence of $\alpha$-coupling constant and the magnitude of Maxwell and {\it dark photon} fields on the adequate effective potentials.
For the both configurations of magnetic fields, i.e., dipolar and parabolic, the qualitative behavior is the same. The bigger value of $\alpha$-coupling
constant we take into account, the larger values of the effective potentials are achieved. As far as the magnitude of the gauge fields is concerned,
for the case when 
$B^{(B)}_0 > B^{(F)}_0$, we have 
the tendency that
the bigger distances among $U_{eff}(r)$ dedicated to the adequate value of $\alpha$, for the larger coupling constant.
On the contrary, if the magnitude of Maxwell field is far more greater than {\it dark photon} one, we achieve very small distances among those curves, devoted to the specific values of $\alpha$ (almost one line which separates for large values of the distance $r$).

In between, one has the situation ( like in Fig. \ref{Fig:11}), for $B^{(F)}_0 = B^{(B)}_0$ when we see the clear separation of the curves among themselves.

\subsection{Current loops around black hole}
 For the completeness of the studies we also pay attention to magnetic fields generated by the current loops surrounding black hole. The gauge field potentials yield
 \cite{wal74, pet74}, \cite{ull24},
 \be 
 \tA^\mu = \frac{ B^{(\tF)}}{2} F_k(r)~\xi^\mu_{(\phi)}, \qquad  \tB^\mu = \frac{B^{(\tB)}}{2}F_k(r)~\xi^\mu_{(\phi)},
\label{loop}
\ee
where $k=1,~2$ describes the outside and inside of the adequate current loop, while the combinations of 
magnetic fields connected with the loops in question

\be
B^{(\tF)} = \frac{\sqrt{2 -\alpha}}{2} \Big( B^{(F)} - B^{(B)} \Big), \qquad
B^{(\tB)} = \frac{\sqrt{2 +\alpha}}{2} \Big( B^{(F)} + B^{(B)} \Big),
\label{bbbo}
\ee
where the magnetic field strength near the loop of Maxwell current and {\it dark photon} are given respectively by
\be 
B^{(F)} = \pi r_0^2 \sqrt{1 - f(r)} I^{(F)}, \qquad B^{(B)} = \pi r_0^2 \sqrt{1 - f(r)} I^{(B)},
\label{bbb}
\ee
The magnetic field in question is uniform in the inner region, while it resembles a dipole-like form in the outer region. Namely \cite{pet74}
\ben
F_1(r) &=& \ln f(r) + \frac{2M }{r} \Big( 1 + \frac{M}{r} \Big), ~r > r_0, \\
F_2(r) &=& \ln f(r_0) + \frac{2M }{r_0} \Big( 1 + \frac{M}{r_0} \Big), ~2M < r \le r_0.
\een

Let us have a close look at the form of the effective potential given by the relation (\ref{uffo}), having in mind the exact form of $\te_A$ and $\te_B$, as well as, the fact that
$\tB^{(F,B)}_0$ are replaced by quantities given in relation (\ref{bbbo}).
In the case of $e_d =0$, the effective potential reduce to the one given by (\ref{uff}). Then having in mind relations for $w^{(F)}$ 
and $w^{(B)}$ given by the equation (\ref{wfwb}), where 
now $B^{(F)}$ and $B^{(B)}$ are given by (\ref{bbb}) and the fact that current connected with {\it dark sector} charge $I^{(B)}$, is equal to zero, we shall obtain
the same conclusions as in the ordinary Maxwell case \cite{ull24}.

However the example in question can be implemented as the 'possible observational experiment' of the {\it dark photon} current existence.
If observations show the different than Maxwell field results, it will be a strong suspicion that one observes 'new physics', possibly related to the {\it dark photon}
version of the {\it hidden sector}.

\section{Conclusions}

In this paper we have considered motion of magnetized and charged particles
in the spacetime of a static spherically symmetric weakly magnetized black hole. We take into account the Einstein-Maxwell gravity with
{\it dark matter} sector, using the so-called {\it dark photon} theory, where for the {\it invisible sector} is responsible auxiliary $U(1)$-gauge field coupled to the ordinary Maxwell one.
The weakly magnetized black hole solution has been found by means of Wald's procedure \cite{wal74}, utilizing the formal resemblance of equations of motion
for Maxwell-{\it dark photon} fields (in Lorentz gauge) with the the relations for Killing vector fields.

Using Hamilton-Jacobi equations for magnetized particle and restricting our attention to the linear terms (weak magnetic field assumption), one can show
that the magnetic coupling parameter $\beta$, responsible for the strength of external magnetic effects, is influenced by {\it dark matter sector} magnetic field and
$\alpha$-coupling constant. These factors also effect the energy of the particle $\cE$ and its angular momentum $l$.

The inspection of the particle collision process in the nearby of the aforementioned weakly magnetized  black hole was also taken into account. Namely,
we have considered
magnetized/charged particle with energies $\cE_i$, angular momenta $l_i$, $i =1,~2,$ which motions have been subject to magnetic $\beta_{(F,B)}$ and charge $w^{(F,B)}$ parameters.
One examines the repulsive and attractive Lorentz force pointed respectively outward and toward spherically symmetric black hole.
 
It happens that
for the neutral and magnetized particle collision, one achieves the description which is qualitatively the same as for the ordinary Schwarzschild black hole spacetime,
but quantitatively the values of $\cE$ and $l$ will depend on {\it dark matter sector}.

For the case of charged and magnetized particles, having the same energy and 
charge due to Maxwell and {\it dark photon } fields, with the magnetic parameter is equal to $0.5$, we studied the 
 dependence of $\cE_{CM}$ on $r$, for various values of $\alpha$-coupling constant. Due to the very small value of coupling constant,
 we consider the $\alpha$-values to have the clear qualitative view of what happens if the coupling constant of {\it visible} and {\it dark sectors} varies.
It turns out that the larger $\alpha$ one takes into account the bigger value of 
$\cE_{CM}$ we achieve. 

The most interesting case one encounters during examination of
particles ($l_1 = -l_2$) under the assumption that one has greater concentration of {\it dark matter}, where $w^{(B)}> w^{(F)}$.
The bunch of curves, for different values of $\alpha$-coupling constant, cuts in one point  $r_c$ (isosbestic point),
For the distances before and after it the results are quantitatively different, i.e.,
for the distance $r > r_c$ one achieves the larger $\cE_{CM}$ for bigger values of coupling constant, but for $r< r_c$ the situation is opposite.
Consequently, for the high values of the Maxwell fields, there are no difference among the curves connected with different values of $\alpha$-coupling constant.

On the other hand,
the neutral and magnetized particle collision description is qualitatively the same as for the ordinary Schwarzschild black hole spacetime.
However, quantitatively we have the different values of $\cE$ and $l$, because of the fact that magnetic field stemming from {\it dark matter sector} are considered.

Moreover the studies of charged particles in magnetic field configurations (homogeneous, dipolar and parabolic) reveal several interesting features:\\
1. the bigger value of the $\alpha$-coupling constant one implies, the higher value of $U_{eff}$ is achieved,\\
2. the bigger value of {\it dark photon} magnetic field is consider, the larger separation distances among curves devoted to the adequate values of $\alpha$ we get,\\
3. if the Maxwell magnetic field is greater than {\it dark photon} one, we obtain 
$U_{eff}$ curve for specific value of $\alpha$, with very small separation distances (almost one curve), which divides into several one for large distances from
black hole event horizon,\\
4. for the case when $B^{(F)}_0 = B^{(B)}_0$, the situation is in between of those described above, and has the clear separation among the curves, as well as,
the above tendencies are maintained.

Because of the fact that we suppose that $e_d=0$, the {\it dark photon} current loops do not influence on the behavior of the particle motion. However
this fact can serve as the 'possible observational experiment' , i.e.,
if future observations show the different than for Maxwell field result, it will be a strong suspicion that 'new physics' bounded with the {\it dark photon}
version of the {\it hidden sector} is spotted.

\acknowledgments
M. R. and P. V. were partially supported by Grant No. 2022/45/B/ST2/00013 of the National Science Center, Poland.
The authors are grateful to the anonymous referee for the constructive comments and remarks, which helped to improve the manuscript.





\begin{thebibliography}{99}

%
\def\cmp#1#2#3#4{\emph{#4}, \emph{ Commun. Math. Phys.} {\bf #1} (#3) #2}
\def\lmp#1#2#3#4{\emph{#4}, \emph{ Lett. Math. Phys.} {\bf #1} (#3) #2}
\def\hpa#1#2#3#4{\emph{#4}, \emph{ Hell. Phys. Acta} {\bf #1} (#3) #2}
\def\grg#1#2#3#4{\emph{#4}, \emph{ Gen. Rel. Grav.} {\bf #1} (#3) #2}
\def\pr#1#2#3#4{\emph{#4}, \emph{ Phys. Rev.} {\bf #1} (#3) #2}
\def\prl#1#2#3#4{\emph{#4}, \emph{ Phys. Rev. Lett.} {\bf #1}, #2 (#3)}
\def\prd#1#2#3#4{\emph{#4}, \emph{ Phys. Rev. D} {\bf #1}, #2 (#3)}

\def\prb#1#2#3#4{\emph{#4}, \emph{ Phys. Rev. B} {\bf #1}, #2 (#3) }
\def\prx#1#2#3#4{\emph{#4}, \emph{ Phys. Rev. X} {\bf #1} (#3) #2}
\def\pl#1#2#3#4{\emph{#4}, \emph{ Phys. Lett.} {\bf #1} (#3) #2}
\def\pla#1#2#3#4{\emph{#4}, \emph{ Phys. Lett. A} {\bf #1} (#3) #2 }
\def\plb#1#2#3#4{\emph{#4}, \emph{ Phys. Lett. B} {\bf #1}, #2 (#3)}
\def\prep#1#2#3#4{\emph{#4}, \emph{ Phys. Reports} {\bf #1}, #2 (#3)}
\def\phys#1#2#3#4{\emph{#4}, \emph{ Physica} {\bf #1} (#3) #2}
\def\jcp#1#2#3#4{\emph{#4}, \emph{ J. Comput. Phys.} {\bf #1} (#3) #2}
\def\jmp#1#2#3#4{\emph{#4}, \emph{ J. Math. Phys.} {\bf #1} (#3) #2}
\def\jpm#1#2#3#4{\emph{#4}, \emph{ J. Phys. A: Math. Gen.} {\bf #1} (#3) #2}
\def\cpr#1#2#3#4{\emph{#4}, \emph{ Computer Phys. Rept.} {\bf #1} (#3) #2}
\def\cqg#1#2#3#4{\emph{#4}, \emph{ Class. Quant. Grav.} {\bf #1} (#3) #2}
\def\cma#1#2#3#4{\emph{#4}, \emph{ Computers Math. Applic.} {\bf #1} (#3) #2}
\def\mc#1#2#3#4{\emph{#4}, \emph{ Math. Compt.} {\bf #1} (#3) #2}
\def\apj#1#2#3#4{\emph{#4}, \emph{ Astrophys. J.} {\bf #1} (#3) #2}
\def\apjs#1#2#3#4{\emph{#4}, \emph{ Astrophys. J. Suppl.} {\bf #1} (#3) #2}
\def\apjl#1#2#3#4{\emph{#4}, \emph{ Astrophys. J. Lett.} {\bf #1} (#3) #2}
\def\acta#1#2#3#4{\emph{#4}, \emph{ Acta Astronomica} {\bf #1} (#3) #2}
\def\apl#1#2#3#4{\emph{#4}, \emph{ Ann. Physik. (Leipzig)} {\bf #1} (#3) #2}
\def\amjp#1#2#3#4{\emph{#4}, \emph{Am. J. Phys.} {\bf #1} (#3) #2}
\def\anp#1#2#3#4{\emph{#4}, \emph{ Ann. Phys.} {\bf #1} (#3) #2}
\def\sa#1#2#3#4{\emph{#4}, \emph{ Sov. Astro.} {\bf #1} (#3) #2}
\def\sia#1#2#3#4{\emph{#4}, \emph{ SIAM J. Sci. Statist. Comput.} {\bf #1} (#3) #2}
\def\aa#1#2#3#4{\emph{#4}, \emph{ Astron. Astrophys.} {\bf #1} (#3) #2}
\def\mnras#1#2#3#4{\emph{#4}, \emph{ Mon. Not. R. Astr. Soc.} {\bf #1} (#3) #2}
\def\npb#1#2#3#4{\emph{#4}, \emph{ Nucl. Phys. B} {\bf #1}, #2 (#3)}
\def\npa#1#2#3#4{\emph{#4}, \emph{ Nucl. Phys. A} {\bf #1} (#3) #2}

\def\prsla#1#2#3#4{\emph{#4}, \emph{ Proc. R. Soc. London, Ser. A} {\bf #1} (#3) #2}
\def\jhep#1#2#3#4{\emph{#4}, \emph{ JHEP} {\bf #1} (#2) #3}
\def\jcap#1#2#3#4{\emph{#4}, \emph{ JCAP} {\bf #1} (#2) #3}

\def\nuca#1#2#3#4{\emph{#4}, \emph{ Nuovo Cimento A } {\bf #1} (#3) #2}
\def\nucb#1#2#3#4{\emph{#4}, \emph{ Nuovo Cimento B } {\bf #1} (#3) #2}
\def\ijmp#1#2#3#4{\emph{#4}, \emph{ Int. J. Mod. Phys. D} {\bf #1} (#3) #2}
\def\atmp#1#2#3#4{\emph{#4}, \emph{ Adv. Theor. Math. Phys.} {\bf #1} (#3) #2}
\def\ptps#1#2#3#4{\emph{#4}, \emph{ Prog. Theor. Phys. Suppl.} {\bf #1} (#3) #2}
\def\ptp#1#2#3#4{\emph{#4}, \emph{ Prog. Theor. Phys.} {\bf #1} (#3) #2}
\def\lmp#1#2#3#4{\emph{#4}, \emph{ Lett. Math. Phys.} {\bf #1} (#3) #2}
\def\cpam#1#2#3#4{\emph{#4}, \emph{ Comm. Pure Appl. Math.}  {\bf #1} (#3) #2}
\def\adv#1#2#3#4{\emph{#4}, \emph{ Adv. Phys.}  {\bf #1} (#3) #2}
\def\zh#1#2#3#4{\emph{#4}, \emph{ Zh. Eksp. Teor. Fiz.}  {\bf #1} (#3) #2}
\def\mplb#1#2#3#4{\emph{#4}, \emph{ Mod. Phys. Lett. B} {\bf #1}, #2 (#3)}
\def\mpla#1#2#3#4{\emph{#4}, \emph{ Mod. Phys. Lett. A} {\bf #1}, #2 (#3)}


\def\jams#1#2#3#4{\emph{#4}, \emph{ J. Austral. Math. Soc. B} {\bf #1} (#3) #2}
\def\appa#1#2#3#4{\emph{#4}, \emph{ Acta Phys. Polonica A} {\bf #1} (#3) #2}
\def\appb#1#2#3#4{\emph{#4}, \emph{ Acta Phys. Polonica B} {\bf #1} (#3) #2}

\def\nat#1#2#3#4{\emph{#4}, \emph{Nature} {\bf #1} #2 (#3)}
\def\natcom#1#2#3#4{\emph{#4}, \emph{Nature Commun.} {\bf #1} (#3) #2}
\def\natphys#1#2#3#4{\emph{#4}, \emph{Nature Physics} {\bf #1} (#3) #2}
\def\natmat#1#2#3#4{\emph{#4}, \emph{Nature Mat.} {\bf #1} (#3) #2}


\def\science#1#2#3#4{\emph{#4}, \emph{Science} {\bf #1} (#3) #2}
\def\sciadv#1#2#3#4{\emph{#4}, \emph{Sci. Adv.} {\bf #1} (#3) #2}

\def\arcmp#1#2#3#4{\emph{#4}, \emph{Annual Rev. of Cond. Matter Physics} {\bf #1} (#3) #2}
\def\zphys#1#2#3#4{\emph{#4}, \emph{Z. Phys.} {\bf #1}, (#3) #2}
\def\ncs#1#2#3#4{\emph{#4}, \emph{Nuovo Cimento Suppl.} {\bf #1} (#3) #2}
\def\physb#1#2#3#4{\emph{#4}, \emph{Physica B} {\bf #1}, (#3) #2}
\def\jpcm#1#2#3#4{\emph{#4}, \emph{J. Phys.: Condens. Matter } {\bf #1} (#3) #2}
\def\pnas#1#2#3#4{\emph{#4}, \emph{Proc. Nat. Academy Sciences} {\bf #1} (#3) #2}
\def\sssr#1#2#3#4{\emph{#4}, \emph{Izv. Akad Nauk SSSR, ser. fiz.} {\bf #1} (#3) #2}
\def\jpg#1#2#3#4{\emph{#4}, \emph{ J. Phys. G} {\bf #1} (#3) #2}
\def\chinpb#1#2#3#4{\emph{#4}, \emph{Chin. Phys. B} {\bf #1} (#3) #2}
\def\njp#1#2#3#4{\emph{#4}, \emph{ New J. Phys.} {\bf #1} (#3) #2}
\def\frontphys#1#2#3#4{\emph{#4}, \emph{ Front. Phys.} {\bf #1} (#3) #2}
\def\epl#1#2#3#4{\emph{#4}, \emph{ EPL} {\bf #1} (#3) #2}
\def\rmp#1#2#3#4{\emph{#4}, \emph{ Rev. Mod. Phys.} {\bf #1}, #2 (#3)}
\def\rpp#1#2#3#4{\emph{#4}, \emph{ Rep. Prog. Phys.} {\bf #1}, #2 (#3)}

\def\hepph#1#2{{ hep-ph }{#1} (#2)}
\def\arxiv#1#2#3{\emph{#3},{ arXiv }{#1} (#2)}
\def\hepth#1#2{{ hep-th }{#1} (#2)}
\def\grqc#1#2{{ gr-qc }{#1} (#2)}
\def\ibid#1#2#3#4{\emph{#4}, {\it ibid.} {\bf #1} (#3) #2}
\def\conphy#1#2#3#4{\emph{#4}, \emph{Contemporary Physics} {\bf #1}, (#3) #2}
\def\ppnp#1#2#3#4{\emph{#4}, \emph{ Prog. Part. Nucl. Phys} {\bf #1} (#3) #2}
\def\arnps#1#2#3#4{\emph{#4}, \emph{ Annu. Rev. Nucl. Part. Sci.} {\bf #1} (#3) #2}
\def\ijmpa#1#2#3#4{\emph{#4}, \emph{ Int. J. Mod. Phys. A} {\bf #1}, #2 (#3)}
\def\jams#1#2#3#4{\emph{#4}, \emph{ J. Austral. Math. Soc. B} {\bf #1} (#3) #2}
\def\appa#1#2#3#4{\emph{#4}, \emph{ Acta Phys. Polonica A} {\bf #1}, (#3) #2}
\def\nat#1#2#3#4{\emph{#4}, \emph{Nature} {\bf #1}, (#3) #2}
\def\science#1#2#3#4{\emph{#4}, \emph{Science} {\bf #1}, (#3) #2}
\def\arcmp#1#2#3#4{\emph{#4}, \emph{Annual Rev. of Cond. Matter Physics} {\bf #1}, (#3) #2}
\def\jcap#1#2#3#4{\emph{#4}, \emph{JCAP} {\bf #1}, (#3) #2}
\def\conphy#1#2#3#4{\emph{#4}, \emph{Contemporary Physics} {\bf #1}, (#3) #2}
\def\ptps#1#2#3#4{\emph{#4}, \emph{ Prog. Theor. Phys. Suppl.} {\bf #1} (#3) #2}
\def\ptp#1#2#3#4{\emph{#4}, \emph{ Prog. Theor. Phys.} {\bf #1} (#3) #2}
\def\apjsup#1#2#3#4{\emph{#4}, \emph{ Astrophys. J. Suppl. Ser.} {\bf #1} (#3) #2}
\def\eurphysjc#1#2#3#4{\emph{#4}, \emph{ Eur. Phys. J.  C} {\bf #1}, #2 (#3)}
\def\njp#1#2#3#4{\emph{#4}, \emph{ New J. Phys. } {\bf #1} (#3) #2}
\def\eurphysjplus#1#2#3#4{\emph{#4}, \emph{ Eur. Phys. J.  Plus} {\bf #1}, #2 (#3)}
%
\def\hepph#1#2{{ hep-ph }{#1} (#2)}
\def\hepth#1#2{{ hep-th }{#1} (#2)}
\def\astroph#1#2{{ astro-ph }{#1} (#2)}
\def\grqc#1#2{{ gr-qc }{#1} (#2)}
\def\ibid#1#2#3#4{\emph{#4}, {\it ibid.} {\bf #1} (#3) #2}

\def\contp#1#2#3#4{\emph{#4}, \emph{ Contemporary Physics} {\bf #1}, #2 (#3)}
\def\physdarkun#1#2#3#4{\emph{#4}, \emph{ Phys. of Dark Universe } {\bf #1}, #2 (#3)}
\def\astrsc#1#2#3#4{\emph{#4}, \emph{Astrophys. Space Sci.} {\bf #1}, #2 (#3)}

\def\epjc#1#2#3#4{\emph{#4}, \emph{ Eur. Phys. J. C} {\bf #1} #2 (#3) }
\def\revphys#1#2#3#4{\emph{#4}, \emph{Reviews in Phys.} {\bf #1} #2 (#3) }







%




\bibitem{planck}
Planck collaboration, P.A.R.  Ade et al., \aa{571}{A16}{2014}{Planck 2013 results. XVI. Cosmological parameters}.

\bibitem{ber18a}
G. Bertone and D. Hoope, \rmp{90}{507}{2018}{History of dark matter}.
\bibitem{ber18b}
G. Bertone and T.M.P. Tait, \nat{562}{51}{2018}{A new era in the search for dark matter}.

\bibitem{sma23}
C. Smarra et al., \prl{131}{171001}{2023}{Second data release from the European pulsar timing array:
challenging the ultralight dark matter paradigm}.

\bibitem{hol86}
B. Holdom, \plb{166}{196}{1986}{Two $U(1)$'s and $\ep$ charge shifts}.

\bibitem{ach16}
B.S. Acharya, S.A.R. Ellis, G.L. Kane, B.D. Nelson, and M.J. Perry, \prl{117}{181802}{2016}{Lightest Visible-Sector Supersymmetric Particle is Likely Unstable}.

\bibitem{gra16}
P.W. Graham, J. Mardon, and S. Rajendran, \prd{93}{103520}{2016}{Vector dark matter from inflationary fluctuations}
\bibitem{sat22}
T. Sato, F. Takahashi, and M. Yamada, \jcap{08}{2022}{022}{Gravitational production of dark photon dark matter with mass generated by the Higgs mechanism}

\bibitem{axionosc}
P. Agrawal, N. Kitajima, M. Reece, T. Sekiguchi, and F. Takahashi, \plb{801}{135136}{2020}{Relic abundance of dark photon dark matter},\\
 R.T. Co, A. Pierce, Z. Zhang, and Y. Zhao, \prd{99}{075002}{2019}{Dark photon dark matter produced by axion oscillations}, \\
M. Bastero-Gil, J. Santiago, L. Ubaldi, and R. Vega- Morales, \jcap{04}{2019}{015}{Vector dark matter production at the end of inflation}.
\bibitem{dro19}
J.A. Dror, K. Harigaya, and V. Narayan, \prd{99}{035036}{2019}{Parametric resonance production of ultralight vector dark matter}.

\bibitem{reheat}
A. Ahmed, B. Grzadkowski, and A. Socha, \jhep{08}{2020}{059}{Gravitational production of vector dark matter},\\
Y. Ema, K. Nakayama, and Y. Tang, \jhep{09}{2018}{135}{
Production of purely gravitational dark matter}.

\bibitem{cstrings}
 A.J. Long and L.-T. Wang, \prd{99}{063529}{2019}{ Dark photon dark matter from a network of cosmic strings},\\
 N. Kitajima and K. Nakayama, {\it Dark photon dark matter from cosmic strings and gravitational wave background}, \hepth{2212.13573}{2022}.

\bibitem{jea03}
P. Jean {\it et al.}, \aa{407}{L55}{2003}{Early SPI/INTEGRAL measurements of 511 keV line emission from the 4th quadrant of the Galaxy}.
\bibitem{cha08}
J. Chang {\it et al.}, \nat{456}{362}{2008}{An excess of cosmic ray electrons at energies of 300-800 GeV}.
\bibitem{bub14}
E. Bulbul et al., \apj{789}{13}{2014}{Detection of an unidentified emission line in the stacked X-ray spectrum of galaxy clusters}.

\bibitem{fil20}
A. Filippi and M. De Napoli, \revphys{5}{100042}{2020}{Searching in the dark: the hunt for the dark photon}.
\bibitem{ger15}
A. Geringer-Sameth and M.G. Walker, \prl{115}{081101}{2015}{Indication of Gamma-Ray Emission from the Newly Discovered Dwarf Galaxy Reticulum II}.
\bibitem{bod15}
K.K. Boddy and J. Kumar, \prd{92}{023533}{2015}{Indirect detection of {\it dark matter} using MeV-range gamma-rays telescopes}.
\bibitem{til15}
K.Van Tilburg, N. Leefer, L. Bougas, and D. Budker, \prl{115}{011802}{2015}{Search for Ultralight Scalar Dark Matter with Atomic Spectroscopy}.
\bibitem{cha17}
J.H. Chang, R. Essig, and S.D. McDermott, \jhep{01}{2017}{107}{Revisiting Supernova 1987A constraints on dark photons}.
\bibitem{sensei}
M. Crisler et. al. (SENSEI Collaboration), \prl{121}{061803}{2019}{SENSEI: First Direct-Detection Constraints on Sub-GeV Dark Matter from a Surface Run}
\bibitem{lee14}
J.P. Lees et al., \prl{113}{201801}{2014}{Search for a Dark Photon in $e^+ e^-$ Collisions at BABAR}.
\bibitem{dav11}
M. Davier et al., \epjc{71}{1515}{2011}{Reevaluation of the hadronic contributions to the muon g-2 and to $\alpha(M^2_z)$}. 

\bibitem{cap21}
A. Caputo, A.J. Miller, C.A.J. O'Hare, and E. Vitagliano, \prd{104}{095029}{2021}{Dark photon limits: A handbook}.


\bibitem{lu22}
B.-Q. Lu and C.-W. Chiang, \prd{105}{123017}{2022}{Probing dark gauge boson with observations from neutron stars}.
\bibitem{rom23}
A. Romanenko et al., \prl{130}{261801}{2023}{Search for dark photons with superconducting radio frequency cavities}.

\bibitem{fil23}
M. Filzinger, S. D\"orscher, R. Lange, J. Klose, M. Steinel, E. Benkler, E. Peik, C. Lisdat, and N. Huntemann, \prl{130}{253001}{2023}
{Improved Limits on the Coupling of Ultralight Bosonic Dark Matter to Photons from Optical Atomic Clock Comparisons}.
\bibitem{ram23}
K. Ramanathan, N. Klimovich, R. Basu Thakur, B.H. Eom, H.G. Leduc, S. Shu, A.D. Beyer, and P.K. Day,
\prl{130}{231001}{2023}{Wideband Direct Detection Constraints on Hidden Photon Dark Matter with the QUALIPHIDE Experiment}.
\bibitem{kot23}
 S. Kotaka, S. Adachi, R. Fujinaka, S. Honda, H. Nakata, Y. Seino, Y. Sueno, T. Sumida, J. Suzuki, O. Tajima, and S. Takeichi, \prl{130}{071805}
 {2023}{Search for Dark Photon Dark Matter in the Mass Range $74-110\text{ }\text{ }\mathrm{\ensuremath{\mu}}\mathrm{eV}$ with a Cryogenic Millimeter-Wave Receiver}.

\bibitem{kic19}
B. Kiczek and M. Rogatko, \jcap{09}{2019}{049}{Ultra-compact spherically symmetric dark matter charged star}.
\bibitem{kic20}
B. Kiczek and M. Rogatko, \prd{101}{084035}{2020}{Influence of black matter on black scalar hair}.
\bibitem{kic21}
B. Kiczek and M. Rogatko, \prd{103}{124021}{2021}{Axion-like dark matter clouds around rotating black holes}.
\bibitem{kic22}
B. Kiczek and M. Rogatko, \epjc{82}{586}{2022}{Static axion-like dark matter clouds around magnetized rotating wormholes - probe limit case}.
\bibitem{rog23}
M. Rogatko, \prd{108}{064026}{2023}{Uniqueness of dark matter magnetized static black hole spacetime}.
\bibitem{rog24}
M. Rogatko, \prd{109}{104030}{2024}{Dark photon - dark energy stationary axisymmetric black holes}.





\bibitem{abd23}
F. Abdulxamidov, J. Rayimbaev, A. Abdujabbarov, and Z. Stuchlik, \prd{108}{044030}{2023}{Spinning magnetized particles orbiting magnetized Schwarzschild black holes} 


\bibitem{ray23}
J. Rayimbaev, K. F. Dialektopoulos, F. Sarikulov, and A. Abdujabbarov, \eurphysjc{83}{572}{2023}{Quasiperiodic oscillations around hairy black holes in Horndeski gravity}

\bibitem{abd22} F. Abdulxamidov,  C. A. Benavides-Gallego, W.-B. Han, J. Rayimbaev, and A. Abdujabbarov, \prd{106}{024012}{2022}{Spinning test particle motion around a rotating wormhole}
\bibitem{ben21} 
C. A. Benavides-Gallego, W.-B. Han, D. Malafarina, B. Ahmedov, and A. Abdujabbarov, \prd{104}{084024}{2021}{Spinning test particle motion around a traversable wormhole}.

\bibitem{bob23}
M. Boboqambarova, B. Turimov, and A. Abdujabbarov, \mpla{38}{2350071}{2023}{Particle motion around Schwarzschild-MOG black hole}.


\bibitem{fro10}
V. P. Frolov and A. A. Shoom, \prd{82}{084034}{2010}{Motion of charged particles near a weakly magnetized Schwarzschild black hole}.
\bibitem{fro12}
V. P. Frolov, \prd{85}{024020}{2012}{Weakly magnetized black holes as particle accelerators}. 



\bibitem{zah13}
A. M. Al Zahrani, V. P. Frolov, and A. A. Shoom, \prd{87}{084043}{2013}{Critical escape velocity for a charged particle moving around a weakly magnetized Schwarzschild black hole}.

\bibitem{fel03}
F. de Felice and F. Sorge, \cqg{20}{469}{2003}{Magnetized orbits around a Schwarzschild black hole}.

\bibitem{ray16}
J. R. Rayimbaev, \astrsc{361}{288}{2016}{Magnetized particle motion around non-Schwarzschild black hole immersed in an external uniform magnetic field}.

\bibitem{abd20}
A. Abdujabbarov, J. Rayimbaev, B. Turimov, and F. Atamurotov, \physdarkun{30}{100715}{2020}{Dynamics of magnetized particles around 4-D Einstein Gauss-Bonnet black hole}. 


\bibitem{wal74}
R. M. Wald, \prd{10}{1680}{1974}{Black hole in a uniform magnetic field}.

\bibitem{pet74}
J. A. Petterson, \prd{10}{3166}{1974}{Magnetic field of a current loop around a Schwarzschild black hole}.

\bibitem{qi23}
M. Qi, J. Rayimbaev, and B. Ahmedov, \epjc{83}{730}{2023}{Charged particles and quasiperiodic oscillations around magnetized Schwarzschild black hole}.
\bibitem{ull24}
S. Ullah Khan, O. Abdurkhmonov, J. Rayimbaev, S. Ahmedov, Y. Turaev, and S. Muminov, \epjc{84}{650}{2024}{Circular motion and collisions of particles with magnetic dipole moment
and electric charge in dipolar magnetosphere around Schwarzschild black holes}.

\bibitem{mcc24}
F. McCarthy, D. Pirvu, J. Colin Hill, J. Huang, M. C. Johnson, K. K. Rogers, \prl{133}{141003}{2024}{Dark photon limits from patchy screening of the cosmic microwave background}.











\end{thebibliography}
\end{document}